\documentclass[11pt,a4paper,english,amssymb,nofootinbib,superscriptaddress,showpacs]{revtex4}
\usepackage{lmodern}

\usepackage[T1]{fontenc}
\usepackage[latin9]{inputenc}
\usepackage{babel}
\usepackage{amsmath}
\usepackage{amssymb}
\usepackage{esint}
\usepackage[unicode=true,pdfusetitle,
 bookmarks=true,bookmarksnumbered=false,bookmarksopen=false,
 breaklinks=false,pdfborder={0 0 1},backref=false,colorlinks=false]
 {hyperref}

\makeatletter

\pdfpageheight\paperheight
\pdfpagewidth\paperwidth

\@ifundefined{textcolor}{}
{%
 \definecolor{BLACK}{gray}{0}
 \definecolor{WHITE}{gray}{1}
 \definecolor{RED}{rgb}{1,0,0}
 \definecolor{GREEN}{rgb}{0,1,0}
 \definecolor{BLUE}{rgb}{0,0,1}
 \definecolor{CYAN}{cmyk}{1,0,0,0}
 \definecolor{MAGENTA}{cmyk}{0,1,0,0}
 \definecolor{YELLOW}{cmyk}{0,0,1,0}
 }

\usepackage{babel}

\@ifundefined{definecolor}{\usepackage{color}}{}
\usepackage{babel}

\pdfpageheight\paperheight
\pdfpagewidth\paperwidth

\@ifundefined{textcolor}{}{%
 \definecolor{BLACK}{gray}{0}
 \definecolor{WHITE}{gray}{1}
 \definecolor{RED}{rgb}{1,0,0}
 \definecolor{GREEN}{rgb}{0,1,0}
 \definecolor{BLUE}{rgb}{0,0,1}
 \definecolor{CYAN}{cmyk}{1,0,0,0}
 \definecolor{MAGENTA}{cmyk}{0,1,0,0}
 \definecolor{YELLOW}{cmyk}{0,0,1,0}
 }

\@ifundefined{definecolor}{\usepackage{color}}{}
\usepackage{latexsym}\usepackage{bm}

\makeatother

\begin{document}

\title{New Exact Solutions of Quadratic Curvature Gravity}

\author{Metin Gürses}

\email{gurses@fen.bilkent.edu.tr}

\affiliation{{\small Department of Mathematics, Faculty of Sciences}\\
 {\small Bilkent University, 06800 Ankara, Turkey}}

\author{Tahsin Ça\u{g}r\i{} \c{S}i\c{s}man}

\email{tahsin.c.sisman@gmail.com}

\affiliation{Department of Physics,\\
 Middle East Technical University, 06800 Ankara, Turkey}

\author{Bayram Tekin}

\email{btekin@metu.edu.tr}

\affiliation{Department of Physics,\\
 Middle East Technical University, 06800 Ankara, Turkey}

\date{\today}
\begin{abstract}
It is a known fact that the Kerr-Schild type solutions in general
relativity satisfy both exact and linearized Einstein field equations.
We show that this property remains valid also for a special class
of the Kerr-Schild metrics in arbitrary dimensions in generic quadratic
curvature theory. In addition to the AdS-wave (or Siklos) metric which
represents plane waves in an AdS background, we present here a new
exact solution, in this class, to the quadratic gravity in $D$-dimensions
which represents a spherical wave in an AdS background. The solution
is a special case of the Kundt metrics belonging to spacetimes with
constant curvature invariants. 
\end{abstract}

\pacs{04.50.-h, 04.20.Jb, 04.30.-w}

\maketitle
{\small \tableofcontents{}}{\small \par}

\section{Introduction}

Whatever the full UV-finite quantum gravity is, its successful low
energy limit, general relativity (GR), is based on the Riemannian
geometry. In this context finding exact Riemannian spacetimes as solutions
to Einstein's equations (with or without a cosmological constant and/or
sources ) has evolved to be a fine art on its own. There are at least
two books \cite{Stephani,Podolsky} that compile and classify these
spacetimes, discuss their physical interpretations and present techniques
of finding solutions. Like any other low energy theory, GR is expected
to receive corrections at high energies built on more powers of curvature
starting with the quadratic gravity which is the subject of this work.
Even though much has been studied in quadratic gravity theories, compared
to Einstein's theory very little is known about the exact solutions
in generic $D$-dimensions ($D=3$ and $D=4$ are somewhat special
as we shall discuss below). There has been a revival of interest in
quadratic gravity theories because of three recent enticing developments:
a specific quadratic gravity model in $\left(2+1\right)$ dimensions
dubbed as the new massive gravity (NMG) \cite{BHT-PRL} provided the
first example of a parity invariant nonlinear unitary theory with
massive gravitons in its perturbative spectrum. The second development
was the introduction of {}``critical gravity'' \cite{LuPope,DeserLiu}
built from the Ricci scalar, the square of the Weyl tensor and a tuned
cosmological constant that has the same perturbative spectrum as the
Einstein's theory with an improved UV behavior. The third one is the
observation that with Neumann boundary conditions on the metric non-Einstein
solutions of the conformal gravity are eliminated and the theory reduces
to the cosmological Einstein's gravity in $D=4$ dimensions \cite{Maldacena}.
All these developments in quadratic curvature gravity theories prompted
us to study systematically some exact solutions of these theories.

In this work, we will present special Kundt type radiating solutions
\cite{Kundt,ColeyHervik} to quadratic gravity theories in generic
$D$ dimensions. This will be a $D$-dimensional generalization of
the works in three dimensions \cite{chakhad-2009,Aliev-PLB} %
\footnote{In \cite{gurses-sisman-tekin-2011}, for $D=3$, Kundt type solutions
of NMG \cite{chakhad-2009,Aliev-PLB} are used to generate solutions
of $f\left(R_{\mu\nu}\right)$ theories which naturally includes the
generic quadratic curvature theory.%
}. Subclasses of Kundt metrics in various forms have also been studied
as solutions of topologically massive gravity \cite{DJT-PRL,DJT-Annals}
in \cite{Nutku,Gurses,Chow-Classify,chakhad-2009,Chow-Kundt,Gurses-Killing,Aliev-PRL,Aliev-PLB,Aliev-PRD}.
In $D$-dimensions, the AdS-wave metric (also called the Siklos metric
\cite{Siklos,Chamblin}) which is a Kundt metric of Type N with a
cosmological constant was shown to be a solution of the quadratic
curvature theories \cite{Gullu_Gurses} generalizing the result in
$D=3$ \cite{Giribet}. All Einstein spacetimes of Type N solve this
theory exactly in $D$ dimensions \cite{Pravda,Malek}. It is a known
fact that in $D=4$ all Einstein spaces solve quadratic theory exactly.
Critical quadratic gravity has genuinely new solutions with asymptotically
non-AdS geometry that has Logarithmic behavior in Poincare and global
coordinates \cite{Alishah,Gullu_Gurses}. It is important here to
note that the works of Coley \emph{et al.} \cite{Kundt,ColeyHervik,coley-hervik-pelavas-2006,Hervik,coley,coley-hervik-pelavas-2008,coley-hervik-pelavas-2009}
on the classification of pseudo-Riemannian spacetimes, on spacetimes
with constant invariants (CSI) and on Kundt spacetimes in general
relativity have attracted many researchers \cite{chakhad-2009,Chow-Classify,Chow-Kundt,Fuster}
to use them in higher order curvature theories in arbitrary dimensions.
Another important point is that all those metrics solving higher order
curvature theories belong to both Kundt and Kerr-Schild classes, \cite{KerrSchild,GursesGursey,Stephani}.

The layout of the paper is as follows: In the next section, we discuss
the Kerr-Schild class of metrics in AdS backgrounds possessing some
special properties. These properties are so effective that some tensorial
quantities, like Ricci and Riemann tensors become linear in the metric
{}``perturbation'' around the AdS background. In the third section,
we show that the full quadratic gravity field equations reduce to
a fourth order linear partial differential equation. We give a new
exact solution which we call a spherical-AdS wave that has asymptotically
AdS and asymptotically non-AdS; i.e. Log mode behavior just like the
previously found AdS wave. In Section IV, we show that the same class
solve the linearized quadratic gravity field equations. We delegate
the details of the computations to the Appendices.

\section{A Special Class of Kerr-Schild Metrics}

Let us take a $D$-dimensional metric in the Kerr-Schild form \cite{KerrSchild,GursesGursey}
\begin{equation}
g_{\mu\nu}=\bar{g}_{\mu\nu}+2V\lambda_{\mu}\lambda_{\nu},\label{metric1}
\end{equation}
 where $\bar{g}_{\mu\nu}$ is the metric of the AdS space and $V$
is a function of spacetime (see \cite{Anabalon} for some properties
of the Kerr-Schild metrics with generic backgrounds and see also \cite{Pravda-KSwithAdS,Malek}
with an AdS background). The vector $\lambda^{\mu}=g^{\mu\nu}\lambda_{\nu}$
is assumed to be null; i.e. $\lambda_{\mu}\lambda^{\mu}=g_{\mu\nu}\lambda^{\mu}\lambda^{\nu}=0$
and geodesic $\lambda^{\mu}\nabla_{\mu}\lambda_{\rho}=0$. These two
assumptions imply 
\[
\bar{g}_{\mu\nu}\lambda^{\mu}\lambda^{\nu}=0,\qquad\lambda_{\mu}=\bar{g}_{\mu\nu}\lambda^{\nu},\qquad\lambda^{\mu}\bar{\nabla}_{\mu}\lambda_{\rho}=0,
\]
 where the barred covariant derivative is with respect to $\bar{g}_{\mu\nu}$.
The inverse metric can be found as 
\begin{equation}
g^{\mu\nu}=\bar{g}^{\mu\nu}-2V\lambda^{\mu}\lambda^{\nu}.
\end{equation}
 Writing the metric in the form (\ref{metric1}) will help us in explicitly
observing the fact that the solutions of the field equations of the
quadratic gravity are also solutions of the linearized field equations
of the theory with $h_{\mu\nu}\equiv2V\lambda_{\mu}\lambda_{\nu}$.
AdS wave or Siklos spacetimes are in this class with the line element
\begin{align}
ds^{2} & =\frac{1}{k^{2}z^{2}}\left(-dt^{2}+dx^{2}+\sum_{m=1}^{D-3}\left(dx^{m}\right)^{2}+dz^{2}\right)+2V\left(t,x,x_{m},z\right)\lambda_{\mu}\lambda_{\nu}dx^{\mu}\otimes dx^{\nu}\nonumber \\
 & =\frac{1}{k^{2}z^{2}}\left(2dudv+\sum_{m=1}^{D-3}\left(dx^{m}\right)^{2}+dz^{2}\right)+2V\left(u,x_{m},z\right)du^{2},\label{metric2}
\end{align}
 where in the second line we have used the null coordinates defined
as $u=\left(x+t\right)/\sqrt{2}$, $v=\left(x-t\right)/\sqrt{2}$
and chosen $\lambda_{\mu}dx^{\mu}=du$ and $\lambda^{\mu}\partial_{\mu}V=0$
that is $V$ does not depend on $v$. The constant $k^{2}$ is related
to the cosmological constant as $-k^{2}=\frac{2\Lambda}{\left(D-1\right)\left(D-2\right)}$.
With these assumptions, $\lambda^{\mu}$ becomes divergence free (non-expanding)
with respect to the full and background metrics namely $\nabla_{\mu}\lambda^{\mu}=\bar{\nabla}_{\mu}\lambda^{\mu}=0$,
and the Ricci scalar turns out to be a constant given as $R=-D(D-1)k^{2}$.
Besides being non-expanding, it is possible to show that $\lambda^{\mu}$
is a shear-free, $\nabla^{\mu}\lambda^{\nu}\nabla_{(\mu}\lambda_{\nu)}=0$,
and non-twisting, $\nabla^{\mu}\lambda^{\nu}\nabla_{[\mu}\lambda_{\nu]}=0$,
vector. As $\lambda_{\mu}$ is a null vector which is non-expanding,
shear-free and non-twisting, AdS-wave is a Kundt spacetime by definition.
Furthermore, the Weyl tensor satisfies the following property

\begin{equation}
C_{\alpha\beta\gamma\sigma}\lambda^{\sigma}=0,\label{typeN}
\end{equation}
 therefore, $\lambda_{\mu}$ is a null direction of the Weyl tensor.
In $D=4$, (\ref{typeN}) is equivalent to the metric being of Type
N %
\footnote{We thank T.~Málek for pointing us that (\ref{typeN}) is not equivalent
to the defining property of Type-N spacetimes for $D>4$.%
}. Note that $\lambda_{\mu}$ is \emph{not} a Killing vector, but $\zeta_{\mu}\equiv\frac{1}{z^{2}}\lambda_{\mu}$
is a null Killing vector. Recently, it was shown that the AdS-wave
metric (\ref{metric2}) solves the quadratic gravity field equations
in $D$-dimensions provided that the function $V$ satisfies a fourth
order linear partial differential equation which was solved in the
most general setting \cite{Gullu_Gurses}.

In this work, we present a new Kundt solution of the quadratic gravity
field equations in $D$-dimensions which is also in the Kerr-Schild
form (\ref{metric1}) as the AdS-wave. The new solution is similar
to the AdS-wave metric in form, but with a different $\lambda_{\mu}$
which dramatically changes the spacetime. To reach the new metric,
let us rewrite the background AdS in the spherical coordinates which
turns the full metric to 
\begin{align}
ds^{2} & =\frac{1}{k^{2}z^{2}}\left[-dt^{2}+\sum_{m=1}^{D-2}\left(dx^{m}\right)^{2}+dz^{2}\right]+2V\lambda_{\mu}\lambda_{\nu}dx^{\mu}\otimes dx^{\nu}\nonumber \\
 & =\frac{1}{k^{2}r^{2}\cos^{2}\theta}\left[-dt^{2}+dr^{2}+r^{2}d\Omega_{D-2}^{2}\right]+2V\lambda_{\mu}\lambda_{\nu}dx^{\mu}\otimes dx^{\nu},\label{eq:AdS_in_spherical_coord}
\end{align}
 where $d\Omega_{D-2}^{2}$ is the metric on the unit sphere in $\left(D-2\right)$-dimensions.
Here, note that since $z>0$, one needs to constrain $\theta$ in
the interval $0\le\theta<\pi/2$. In the spherical coordinates, boundary
of AdS ($z\rightarrow0$) can be reached with the limits $r\rightarrow0$
or/and $\theta\rightarrow\pi/2$. One can define the null coordinates
as $u\equiv\frac{1}{\sqrt{2}}\left(r+t\right)$ and $v\equiv\frac{1}{\sqrt{2}}\left(r-t\right)$,
then (\ref{eq:AdS_in_spherical_coord}) becomes 
\begin{align}
ds^{2} & =\frac{2}{k^{2}\left(u+v\right)^{2}\cos^{2}\theta}\left[2dudv+\frac{\left(u+v\right)^{2}}{2}d\Omega_{D-2}^{2}\right]+2V\left(u,\Omega_{D-2}\right)du^{2},\nonumber \\
 & =\frac{1}{k^{2}\cos^{2}\theta}\left(\frac{4dudv}{\left(u+v\right)^{2}}+d\Omega_{D-2}^{2}\right)+2V(u,\Omega_{D-2})du^{2},\label{eq:Spherical_wave}
\end{align}
 where we have again chosen $\lambda_{\mu}dx^{\mu}=du$ and $\lambda^{\mu}\partial_{\mu}V=0$.
With these assumptions, once again $\nabla_{\mu}\lambda^{\mu}=\bar{\nabla}_{\mu}\lambda^{\mu}=0$.
This metric can be recast in other coordinates as 
\begin{enumerate}
\item Cartesian: 
\begin{equation}
ds^{2}=\frac{1}{k^{2}z^{2}}\left[-dt^{2}+\sum_{m=1}^{D-2}\left(dx^{m}\right)^{2}+dz^{2}\right]+2V\,(\lambda_{\mu}\, dx^{\mu})^{2},\label{adswave}
\end{equation}
 where 
\begin{equation}
\lambda_{\mu}=\left(1,\frac{x^{m}}{r},\frac{z}{r}\right),\qquad m=1,2,\cdots,D-2;~~r^{2}=z^{2}+\sum_{m=1}^{D-2}\left(x^{m}\right)^{2}.\label{lam1}
\end{equation}
 Here, we note that an infinite boost in the $\left(t-x^{1}\right)$-plane
reduces this metric to the AdS wave metric (\ref{metric2}). 
\item \noindent Another form of the above metric can be given as 
\begin{equation}
ds^{2}=dr^{2}+\frac{4\cosh^{2}kr}{k^{2}(u+v)^{2}}dudv+\frac{\sinh^{2}kr}{k^{2}}d\Omega_{D-3}^{2}+2V\left(u,r,\Omega_{D-3}\right)du^{2}.\label{kundtform}
\end{equation}
 This form was given in \cite{coley-hervik-pelavas-2006,Fuster} as
an example of Kundt spacetimes with constant curvature invariants
(CSI). There exists no null Killing vector field of this spacetime.
$D=3$ case of this form of the metric was given \cite{Aliev-PRL,Aliev-PLB}
as the most general Type-N solution of the three-dimensional new massive
gravity (NMG). 
\end{enumerate}
The AdS-wave metric (\ref{metric2}) and the spherical-wave metric
(\ref{eq:Spherical_wave}) have the following (not necessarily independent)
properties which define the Kerr-Schild-Kundt class: 
\begin{enumerate}
\item $\bar{g}_{\mu\nu}$ is the metric of the AdS space, $g_{\mu\nu}=\bar{g}_{\mu\nu}+2V\lambda_{\mu}\lambda_{\nu}$
is the full metric. 
\item The vector $\lambda^{\mu}=g^{\mu\nu}\lambda_{\nu}$ assumed to have
the properties of being null $\lambda_{\mu}\lambda^{\mu}=g_{\mu\nu}\lambda^{\mu}\lambda^{\nu}=0$
and geodesic $\lambda^{\mu}\nabla_{\mu}\lambda_{\rho}=0$. 
\item $V$ is a function of spacetime assumed to satisfy $\lambda^{\mu}\partial_{\mu}V=0$.
This assumption has wonderful implications together with the assumption
$\nabla_{\mu}\lambda^{\mu}=\bar{\nabla}_{\mu}\lambda^{\mu}=0$. With
these assumptions, Riemann and Ricci tensors become linear in $V$
and the scalar curvature becomes constant. 
\item $\nabla_{\mu}\lambda_{\nu}=\lambda_{(\mu}\xi_{\nu)}$, where $\xi^{\mu}\lambda_{\mu}=0$.%
\footnote{Symmetrization is done as usual; i.e. $2A_{(\mu}B_{\nu)}\equiv A_{\mu}B_{\nu}+A_{\nu}B_{\mu}$.%
} 
\item $\lambda_{\mu}$ is non-expanding, $\nabla_{\mu}\lambda^{\mu}=0$,
shear-free, $\nabla^{\mu}\lambda^{\nu}\nabla_{(\mu}\lambda_{\nu)}=0$,
and non-twisting, $\nabla^{\mu}\lambda^{\nu}\nabla_{[\mu}\lambda_{\nu]}=0$
which are implied by the fourth property. Note that one can replace
the full covariant derivative and the metric with the background covariant
derivative and the background metric in these relations, namely $\bar{\nabla}^{\mu}\lambda^{\nu}\bar{\nabla}_{[\mu}\lambda_{\nu]}=0$,
etc. 
\end{enumerate}
These properties are useful in calculating various tensorial quantities.
Here, we note the results of the relevant computations and delegate
some to the Appendix. The Riemann tensor of (\ref{metric1}) after
using some of the properties listed above reduces to 
\begin{equation}
R_{\phantom{\mu}\alpha\nu\beta}^{\mu}=\bar{R}_{\phantom{\mu}\alpha\nu\beta}^{\mu}+\bar{\nabla}_{\nu}\Omega_{\phantom{\mu}\alpha\beta}^{\mu}-\bar{\nabla}_{\beta}\Omega_{\phantom{\mu}\alpha\nu}^{\mu},\label{eq:Riemann-KSK}
\end{equation}
 where
\begin{align}
\bar{\nabla}_{\nu}\Omega_{\phantom{\mu}\alpha\beta}^{\mu}-\bar{\nabla}_{\beta}\Omega_{\phantom{\mu}\alpha\nu}^{\mu}= & 2\lambda_{\alpha}\lambda_{[\nu}\bar{\nabla}_{\beta]}\partial^{\mu}V-2\lambda^{\mu}\lambda_{[\nu}\bar{\nabla}_{\beta]}\partial_{\alpha}V\nonumber \\
 & +\lambda_{[\nu}\xi_{\beta]}\left(\lambda_{\alpha}\partial^{\mu}V-\lambda^{\mu}\partial_{\alpha}V+\lambda_{\alpha}\xi^{\mu}V\right)\nonumber \\
 & +\left(\lambda_{\alpha}\xi^{\mu}-\lambda^{\mu}\xi_{\alpha}\right)\lambda_{[\nu}\partial_{\beta]}V\nonumber \\
 & +2V\lambda^{\mu}\left(\lambda_{\alpha}\bar{\nabla}_{[\nu}\xi_{\beta]}-\lambda_{[\nu}\bar{\nabla}_{\beta]}\xi_{\alpha}\right),
\end{align}
 where the background part reads $\bar{R}_{\mu\alpha\nu\beta}=-k^{2}\left(\bar{g}_{\mu\nu}\bar{g}_{\alpha\beta}-\bar{g}_{\mu\beta}\bar{g}_{\alpha\nu}\right)$
and the remaining part is linear in $V$. The property (\ref{typeN})
leads to 
\begin{equation}
R_{\phantom{\rho}\mu\nu\alpha}^{\rho}\lambda_{\rho}=\frac{R}{D\left(D-1\right)}\left(\lambda_{\alpha}\, g_{\mu\nu}-\lambda_{\nu}\, g_{\mu\alpha}\right).
\end{equation}
 For the class of Kerr-Schild-Kundt metrics, the Ricci tensor follows
from (\ref{eq:Riemann-KSK}) as

\begin{equation}
R_{\mu\nu}=-\left(D-1\right)k^{2}g_{\mu\nu}-\rho\lambda_{\mu}\lambda_{\nu},\label{ein}
\end{equation}
 where 
\begin{equation}
\rho\equiv\bar{\square}V+2\xi_{\mu}\partial^{\mu}V+\frac{1}{2}V\xi_{\mu}\xi^{\mu}-2Vk^{2}\left(D-2\right).\label{eq:rho}
\end{equation}
 where $\bar{\square}\equiv\bar{\nabla}^{\rho}\bar{\nabla}_{\rho}$
and $\lambda^{\mu}\partial_{\mu}\rho=0$ and the Ricci scalar is $R=-D(D-1)\, k^{2}$.
It is amusing to see that the metric solves the cosmological Einstein
equations in the presence of a null fluid in all dimensions as long
as $T_{\mu\nu}=\rho\lambda_{\mu}\lambda_{\nu}$, but our task is to
show that the same metric solves the vacuum field equations of the
quadratic gravity.

Using the properties listed above of the new metric we find the following
tensors that we shall need in the field equations of the most general
quadratic gravity; 
\begin{equation}
\square R_{\mu\nu}=-\bar{\square}\left(\rho\lambda_{\mu}\lambda_{\nu}\right),\label{eq:Box_Ricci_form1}
\end{equation}
 or in another form 
\begin{equation}
\square R_{\mu\nu}=-\lambda_{\mu}\lambda_{\nu}\left(\bar{\square}\rho+2\xi_{\sigma}\partial^{\sigma}\rho+\frac{1}{2}\rho\xi_{\sigma}\xi^{\sigma}-2\rho k^{2}\left(D-1\right)\right),\label{eq:Box_Ricci_form2}
\end{equation}
 and 
\begin{equation}
R_{\mu}^{\rho}R_{\rho\nu}=\left(D-1\right)^{2}k^{4}g_{\mu\nu}+2\left(D-1\right)k^{2}\rho\lambda_{\mu}\lambda_{\nu},\label{eq:Ricci2}
\end{equation}
 
\begin{equation}
R_{\mu\alpha\nu\beta}R^{\alpha\beta}=\left(D-1\right)^{2}k^{4}g_{\mu\nu}+\left(D-2\right)k^{2}\rho\lambda_{\mu}\lambda_{\nu},\label{eq:RiemRic}
\end{equation}
 
\begin{equation}
R_{\mu\alpha\beta\gamma}R_{\nu}^{\phantom{\nu}\alpha\beta\gamma}=2(D-1)k^{4}g_{\mu\nu}+4k^{2}\rho\lambda_{\mu}\lambda_{\nu}.\label{eq:Riem2}
\end{equation}

\section{A New Solution of the Quadratic Gravity}

\noindent The action of the quadratic gravity is 
\begin{eqnarray}
I & = & \int d^{D}x\,\sqrt{-g}\left[\frac{1}{\kappa}\left(R-2\Lambda_{0}\right)+\alpha R^{2}+\beta R_{\mu\nu}^{^{2}}+\gamma\left(R_{\mu\nu\sigma\rho}^{2}-4R_{\mu\nu}^{2}+R^{2}\right)\right].\label{eq:Quadratic_action}
\end{eqnarray}
 The (source-free) field equations were given in \cite{DeserTekin,Gullu_Tekin}
as 
\begin{align}
\frac{1}{\kappa}\left(R_{\mu\nu}-\frac{1}{2}g_{\mu\nu}R+\Lambda_{0}g_{\mu\nu}\right)+2\alpha R\left(R_{\mu\nu}-\frac{1}{4}g_{\mu\nu}R\right)+\left(2\alpha+\beta\right)\left(g_{\mu\nu}\square-\nabla_{\mu}\nabla_{\nu}\right)R\nonumber \\
+2\gamma\left[RR_{\mu\nu}-2R_{\mu\sigma\nu\rho}R^{\sigma\rho}+R_{\mu\sigma\rho\tau}R_{\nu}^{\phantom{\nu}\sigma\rho\tau}-2R_{\mu\sigma}R_{\nu}^{\phantom{\nu}\sigma}-\frac{1}{4}g_{\mu\nu}\left(R_{\tau\lambda\sigma\rho}^{2}-4R_{\sigma\rho}^{2}+R^{2}\right)\right]\nonumber \\
+\beta\square\left(R_{\mu\nu}-\frac{1}{2}g_{\mu\nu}R\right)+2\beta\left(R_{\mu\sigma\nu\rho}-\frac{1}{4}g_{\mu\nu}R_{\sigma\rho}\right)R^{\sigma\rho} & =0.\label{fieldequations}
\end{align}
 Using (\ref{ein}-\ref{eq:Riem2}) in (\ref{fieldequations}), one
obtains 
\begin{equation}
\frac{\Lambda-\Lambda_{0}}{2\kappa}+f\Lambda^{2}=0,\qquad\Lambda\equiv-\frac{\left(D-1\right)\left(D-2\right)}{2}k^{2},\quad f\equiv\left(D\alpha+\beta\right)\frac{\left(D-4\right)}{\left(D-2\right)^{2}}+\gamma\frac{\left(D-3\right)\left(D-4\right)}{\left(D-1\right)\left(D-2\right)},\label{quadratic}
\end{equation}
 as a trace equation, and the remaining traceless equation is a fourth
order equation, 
\begin{equation}
\left(\beta\bar{\square}+c\right)\left(\rho\lambda_{\mu}\lambda_{\nu}\right)=0,\label{eq:Lambda_eqn_from_lin_eom}
\end{equation}
 where 
\begin{equation}
c\equiv\frac{1}{\kappa}+\frac{4\Lambda D}{D-2}\alpha+\frac{4\Lambda}{D-1}\beta+\frac{4\Lambda\left(D-3\right)\left(D-4\right)}{\left(D-1\right)\left(D-2\right)}\gamma.\label{eq:c}
\end{equation}

As noted before, AdS wave \cite{Gullu_Gurses} solves (\ref{eq:Lambda_eqn_from_lin_eom}).
Now, let us find the second solution that is the spherical-AdS-wave
metric (\ref{eq:Spherical_wave}). This can be achieved by obtaining
a fourth order scalar equation on $V$ 
\begin{equation}
\left(\mathcal{O}-M^{2}\right)\mathcal{O}V\left(u,\Omega_{D-2}\right)=0,\label{exactequation}
\end{equation}
 where 
\begin{equation}
M^{2}\equiv-\frac{c}{\beta}+2k^{2},\qquad\mathcal{O}\equiv\bar{\square}-2k^{2}\sin2\theta\partial_{\theta}-2k^{2}\left(D-2-\sin^{2}\theta\right).
\end{equation}
 To reach (\ref{exactequation}), we have calculated $\rho$ for the
spherical-AdS-wave which is $\rho=\mathcal{O}V$. It is important
to notice that there are two different types of solutions to (\ref{exactequation}).
The first type solution is $V=V_{1}+V_{2}$ where $V_{1}$ is a solution
to the quadratic partial differential equation (PDE) 
\begin{equation}
\mathcal{O}V_{1}\left(u,\Omega_{D-2}\right)=0,\label{eq:V_eqn-Einstein_modes}
\end{equation}
 which is also a solution of the cosmological Einstein's theory, ($\rho=0$),
and $V_{2}$ is a solution to again a quadratic PDE 
\begin{equation}
\left(\mathcal{O}-M^{2}\right)V_{2}\left(u,\Omega_{D-2}\right)=0.\label{eq:V_eqn-Massive_modes}
\end{equation}
 As long as $M^{2}\ne0$, $V=V_{1}+V_{2}$ is the most general solution
to the fourth order PDE (\ref{exactequation}). But, when $M^{2}=0$,
then the equation becomes 
\begin{equation}
\mathcal{O}^{2}V\left(u,\Omega_{D-2}\right)=0,\label{eq:V_eqn-Log_modes}
\end{equation}
 and new solutions arise which represent the non-Einstein solutions
of the critical gravity. To get the solutions, let us employ the separation
of variables technique as $V\left(u,\Omega_{D-2}\right)=F\left(u,\theta\right)G\left(u,\Omega_{D-3}\right)$
where $G\left(u,\Omega_{D-3}\right)$ is the function defined on the
$\left(D-3\right)$-dimensional unit sphere. For a scalar function
$\Phi\left(u,\theta,\Omega_{D-3}\right)$, let us calculate $\bar{\nabla}^{\rho}\bar{\nabla}_{\rho}\Phi\left(u,\theta,\Omega_{D-3}\right)$
for the background AdS metric 
\begin{equation}
d\bar{s}^{2}=\frac{4dudv}{k^{2}\cos^{2}\theta\left(u+v\right)^{2}}+\frac{1}{k^{2}\cos^{2}\theta}d\Omega_{D-2}^{2},
\end{equation}
 which corresponds to $V=0$ in (\ref{eq:Spherical_wave}): 
\begin{equation}
\bar{\nabla}^{\rho}\bar{\nabla}_{\rho}\Phi\left(u,\theta,\Omega_{D-3}\right)=2\bar{g}^{vu}\bar{\nabla}_{v}\partial_{u}\Phi\left(u,\theta,\Omega_{D-3}\right)+\bar{g}^{\Omega_{i}\Omega_{i}}\bar{\nabla}_{\Omega_{i}}\partial_{\Omega_{i}}\Phi\left(u,\theta,\Omega_{D-3}\right),\label{eq:Box_phi_comp}
\end{equation}
 where $\Omega_{i}$ represents the angular coordinates on $S^{D-2}$
which includes the $\theta$ direction. Using the results in the Appendix,
the first term yields 
\begin{equation}
2\bar{g}^{vu}\bar{\nabla}_{v}\partial_{u}\Phi\left(u,\theta,\Omega_{D-3}\right)=2k^{2}\sin\theta\cos\theta\partial_{\theta}\Phi\left(u,\theta,\Omega_{D-3}\right).
\end{equation}
 On the other hand, the second term can be written as 
\begin{align}
\bar{g}^{\Omega_{i}\Omega_{i}}\bar{\nabla}_{\Omega_{i}}\partial_{\Omega_{i}}\Phi\left(u,\theta,\Omega_{D-3}\right)= & \bar{g}^{\Omega_{i}\Omega_{i}}\partial_{\Omega_{i}}\partial_{\Omega_{i}}\Phi\left(u,\theta,\Omega_{D-3}\right)-\bar{g}^{\Omega_{i}\Omega_{i}}\bar{\Gamma}_{\Omega_{i}\Omega_{i}}^{\Omega_{j}}\partial_{\Omega_{j}}\Phi\left(u,\theta,\Omega_{D-3}\right)\nonumber \\
 & -\bar{g}^{\Omega_{i}\Omega_{i}}\bar{\Gamma}_{\Omega_{i}\Omega_{i}}^{u}\partial_{u}\Phi\left(u,\theta,\Omega_{D-3}\right),\label{eq:Box_of_S^D-2}
\end{align}
 In the Appendix, it is shown that $\bar{\Gamma}_{\Omega_{i}\Omega_{i}}^{u}=0$;
therefore, the last term vanishes. Then, let us calculate the first
line in (\ref{eq:Box_of_S^D-2}) which corresponds to the box operator
acting on a scalar function with the following metric conformal to
the metric $\eta_{\Omega_{i}\Omega_{j}}$ (not to be confused with
the flat metric) on the round $S^{D-2}$ sphere: 
\begin{equation}
ds^{2}=\frac{1}{k^{2}\cos^{2}\theta}d\Omega_{D-2}^{2}\Rightarrow\bar{g}_{\Omega_{i}\Omega_{j}}=\omega^{-2}\eta_{\Omega_{i}\Omega_{j}},\quad\omega\equiv k\cos\theta.
\end{equation}
 The Christoffel connection of $\bar{g}_{\Omega_{i}\Omega_{j}}$ is
related to the Christoffel connection of $\eta_{\Omega_{i}\Omega_{j}}$
via the usual conformal transformations 
\begin{equation}
\bar{\Gamma}_{\alpha\beta}^{\mu}=\left(\Gamma_{\alpha\beta}^{\mu}\right)_{S^{D-2}}-\delta_{\alpha}^{\mu}\partial_{\beta}\ln\omega-\delta_{\beta}^{\mu}\partial_{\alpha}\ln\omega+\eta_{\alpha\beta}\eta^{\mu\sigma}\partial_{\sigma}\ln\omega,
\end{equation}
 Using this result in $\bar{g}^{\Omega_{i}\Omega_{i}}\bar{\nabla}_{\Omega_{i}}\partial_{\Omega_{i}}\Phi$,
one obtains 
\begin{align}
\bar{g}^{\Omega_{i}\Omega_{i}}\bar{\nabla}_{\Omega_{i}}\partial_{\Omega_{i}}\Phi\left(u,\theta,\Omega_{D-3}\right)= & \omega^{2}\left[\eta^{\Omega_{i}\Omega_{i}}\partial_{\Omega_{i}}\partial_{\Omega_{i}}\Phi\left(u,\theta,\Omega_{D-3}\right)-\eta^{\Omega_{i}\Omega_{i}}\left(\Gamma_{\Omega_{i}\Omega_{i}}^{\Omega_{j}}\right)_{S^{D-2}}\partial_{\Omega_{j}}\Phi\left(u,\theta,\Omega_{D-3}\right)\right]\nonumber \\
 & +\omega^{2}\left[2\eta^{\theta\theta}\delta_{\theta}^{\Omega_{j}}\partial_{\theta}\ln\omega-\eta^{\Omega_{i}\Omega_{i}}\eta_{\Omega_{i}\Omega_{i}}\eta^{\Omega_{j}\theta}\partial_{\theta}\ln\omega\right]\partial_{\Omega_{j}}\Phi\left(u,\theta,\Omega_{D-3}\right),\label{eq:Box_in_S^D-2_LB}
\end{align}
 where the square bracket in the first line is the Laplace-Beltrami
operator on $S^{D-2}$ which can be recursively written as 
\begin{align}
\Delta_{S^{D-2}}\Phi\left(u,\theta,\Omega_{D-3}\right) & =\frac{1}{\sin^{D-3}\theta}\frac{\partial}{\partial\theta}\left(\sin^{D-3}\theta\frac{\partial\Phi\left(u,\theta,\Omega_{D-3}\right)}{\partial\theta}\right)+\frac{1}{\sin^{2}\theta}\Delta_{S^{D-3}}\Phi\left(u,\theta,\Omega_{D-3}\right)\nonumber \\
 & =\left(\frac{\partial^{2}}{\partial\theta^{2}}+\left(D-3\right)\cot\theta\frac{\partial}{\partial\theta}+\frac{1}{\sin^{2}\theta}\Delta_{S^{D-3}}\right)\Phi\left(u,\theta,\Omega_{D-3}\right).\label{eq:Laplace-Beltrami}
\end{align}
 Collecting (\ref{eq:Box_in_S^D-2_LB}) and (\ref{eq:Laplace-Beltrami}),
one arrives at 
\begin{align}
\bar{g}^{\Omega_{i}\Omega_{i}}\bar{\nabla}_{\Omega_{i}}\partial_{\Omega_{i}}\Phi\left(u,\theta,\Omega_{D-3}\right)= & k^{2}\cos^{2}\theta\left(\frac{\partial^{2}}{\partial\theta^{2}}+\left(D-3\right)\cot\theta\frac{\partial}{\partial\theta}+\frac{1}{\sin^{2}\theta}\Delta_{S^{D-3}}\right)\Phi\left(u,\theta,\Omega_{D-3}\right)\nonumber \\
 & +k^{2}\left(D-4\right)\sin\theta\cos\theta\partial_{\theta}\Phi\left(u,\theta,\Omega_{D-3}\right).
\end{align}
 Finally, one has

\begin{align}
\bar{\square}\Phi\left(u,\theta,\Omega_{D-3}\right)= & k^{2}\cos^{2}\theta\frac{\partial^{2}\Phi\left(u,\theta,\Omega_{D-3}\right)}{\partial\theta^{2}}+k^{2}\left[\left(D-3\right)\cot\theta+\sin\theta\cos\theta\right]\frac{\partial\Phi\left(u,\theta,\Omega_{D-3}\right)}{\partial\theta}\nonumber \\
 & +k^{2}\cot^{2}\theta\Delta_{S^{D-3}}\Phi\left(u,\theta,\Omega_{D-3}\right).\label{eq:Box_phi}
\end{align}

This result is sufficient for us to carry out the separation of variables.
Let us first focus on the Einstein modes satisfying (\ref{eq:V_eqn-Einstein_modes}).
Using (\ref{eq:Box_phi}) for $V\left(u,\Omega_{D-2}\right)=F\left(u,\theta\right)G\left(u,\Omega_{D-3}\right)$,
one has two decoupled equations 
\begin{multline}
\cos^{2}\theta\frac{\partial^{2}F\left(u,\theta\right)}{\partial\theta^{2}}+\left[\left(D-3\right)\cot\theta-3\sin\theta\cos\theta\right]\frac{\partial F\left(u,\theta\right)}{\partial\theta}\\
-\left[2\left(D-2-\sin^{2}\theta\right)+a^{2}\left(u\right)\cot^{2}\theta\right]F\left(u,\theta\right)=0,\label{eq:theta_eqn}
\end{multline}
 
\begin{equation}
\left(\Delta_{S^{D-3}}+a^{2}\left(u\right)\right)G\left(u,\Omega_{D-3}\right)=0,\label{eq:Spherical_Laplacian}
\end{equation}
 where $a^{2}$ is an arbitrary function of $u$. Both of these equations
can be solved exactly for $a^{2}\ne0$: (\ref{eq:theta_eqn}) has
a solution in terms of hypergeometric functions and (\ref{eq:Spherical_Laplacian})
in terms of spherical harmonics on $S^{D-3}$ \cite{Higuchi}. Since
the most general solution is not particularly illuminating to depict
here for the sake of simplicity let us concentrate on $D=4$, for
which one has 
\begin{equation}
F\left(u,\theta\right)=\frac{c_{1}\left(u\right)}{a}\left(\tan\frac{\theta}{2}\right)^{a}\sec\theta\left(a+\sec\theta\right)+\frac{c_{2}\left(u\right)}{\left(a^{2}-1\right)}\left(\tan\frac{\theta}{2}\right)^{-a}\sec\theta\left(a-\sec\theta\right),\label{eq:Einstein_mode_theta}
\end{equation}
 
\begin{equation}
G\left(u,\phi\right)=c_{3}\left(u\right)\cos\left(a\phi\right)+c_{4}\left(u\right)\sin\left(a\phi\right).\label{eq:Einstein_mode_spherical}
\end{equation}
 Here, one of the functions $c_{i}\left(u\right)$ can be set to $1$
without loss of generality, if it is not zero. Note that $a=0$ and
$a^{2}=1$ are the special values for which the solutions can be obtained
as: 
\begin{itemize}
\item $D=4$ and $a=0$: 
\begin{equation}
F\left(u,\theta\right)=c_{1}\left(u\right)\sec^{2}\theta+c_{2}\left(u\right)\left(\cos\theta+\log\left[\tan\left(\frac{\theta}{2}\right)\right]\right)\sec^{2}\theta,\label{eq:a=00003D00003D0_theta_eqn_soln}
\end{equation}
 
\begin{equation}
G\left(u,\phi\right)=c_{3}\left(u\right)+c_{4}\left(u\right)\phi.
\end{equation}
 More explicitly, the solution reads 
\begin{equation}
V\left(u,\theta,\phi\right)=\frac{1}{\cos^{2}\theta}\left[1+c_{2}\left(u\right)\left(\cos\theta+\log\left[\tan\left(\frac{\theta}{2}\right)\right]\right)\right]\left(c_{3}\left(u\right)+c_{4}\left(u\right)\phi\right).\label{eq:a=00003D00003D0_V}
\end{equation}
 Let us investigate the near boundary behavior of this metric by defining
$x\equiv\pi/2-\theta$ and finding the asymptotic form for small $x$.
In order to have complete comparison with the AdS-wave boundary behavior,
one needs to expand up to $O\left(x^{4}\right)$ which yields 
\begin{equation}
F\left(u,x\right)\sim\frac{1}{x^{2}}\left[1+\frac{1}{3}x^{2}+c_{2}\left(u\right)x^{3}+O\left(x^{4}\right)\right].
\end{equation}
 Here, the leading order represents the asymptotically AdS spacetime
just like the AdS wave; while the next-to-leading order; i.e. $O\left(1/x\right)$,
shows that the spherical-AdS-wave asymptotes to AdS spacetime more
slowly than the AdS-wave which exactly behaves as 
\begin{equation}
V_{\text{AdS-wave}}\left(u,x\right)=\frac{1}{x^{2}}\left[1+c_{2}\left(u\right)x^{3}\right].\label{eq:AdS-wave_V}
\end{equation}

\item $D=4$ and $a^{2}=1$ is also a simple solution which we depict here:
\begin{equation}
F\left(u,\theta\right)=c_{1}\left(u\right)\sec\theta\tan\theta+c_{2}\left(u\right)\csc\theta\left(\log\left[\tan\left(\frac{\theta}{2}\right)\right]-\sec\theta+{\rm arctanh}\left[\cos\theta\right]\sec^{2}\theta\right),
\end{equation}
 
\begin{equation}
G\left(u,\phi\right)=c_{3}\left(u\right)\cos\left(\phi\right)+c_{4}\left(u\right)\sin\left(\phi\right).
\end{equation}

\end{itemize}
Clearly, the solutions of (\ref{eq:V_eqn-Massive_modes}), which we
call massive modes, have the same functional form as the Einstein
modes in (\ref{eq:Einstein_mode_theta}) and (\ref{eq:Einstein_mode_spherical}).
In order to obtain the massive modes explicitly, the only thing one
should do is to replace $a$ in (\ref{eq:Einstein_mode_theta}) with
$\sqrt{a^{2}+M^{2}}$.

Now, let us focus on the non-Einstein solutions of the $M^{2}=0$
case with the field equation (\ref{eq:V_eqn-Log_modes}) corresponding
to the critical gravity. We are interested in the spherical-wave solutions
which spoil the asymptotically AdS nature of the spacetime. Thus,
in order to study the near-boundary behavior, it is enough to study
the $\theta$ dependence of the metric function $V$ by studying the
square of the operator appearing in the $\theta$-equation (\ref{eq:theta_eqn})
as acting on $V\left(u,\theta\right)$ as 
\begin{equation}
\left[\cos^{2}\theta\frac{\partial^{2}}{\partial\theta^{2}}+\left[\left(D-3\right)\cot\theta-3\sin\theta\cos\theta\right]\frac{\partial}{\partial\theta}-2\left(D-2-\sin^{2}\theta\right)\right]^{2}V\left(u,\theta\right)=0.\label{eq:Critical_field_eqn}
\end{equation}
 Besides the homogeneous solutions (\ref{eq:a=00003D00003D0_theta_eqn_soln}),
the particular solution of the equation 
\begin{multline}
\left[\cos^{2}\theta\frac{\partial^{2}}{\partial\theta^{2}}+\left[\left(D-3\right)\cot\theta-3\sin\theta\cos\theta\right]\frac{\partial}{\partial\theta}-2\left(D-2-\sin^{2}\theta\right)\right]V\left(u,\theta\right)\\
=\frac{1}{\cos^{2}\theta}\left[1+c_{2}\left(u\right)\left(\cos\theta+\log\left[\tan\left(\frac{\theta}{2}\right)\right]\right)\right],
\end{multline}
 also provide a solution to (\ref{eq:Critical_field_eqn}). As the
$1/x^{2}$ part of (\ref{eq:AdS-wave_V}) gives rise to the Log mode
which changes the boundary behavior in the AdS-wave case, one may
expect that $1/\cos^{2}\theta$ part of the homogeneous solution (\ref{eq:a=00003D00003D0_theta_eqn_soln}),
having the same near-boundary behavior, should give rise to the Log
mode of the spherical-AdS wave. This expectation is confirmed by investigating
the asymptotic behavior of the particular solution for the source
with $c_{2}\left(u\right)=0$ which can be found as 
\begin{equation}
V_{p}\left(u,\theta\right)=\frac{\log\left[\tan\theta\right]}{3\cos^{2}\theta}.\label{eq:Log-mode}
\end{equation}
 Again with the definition $x\equiv\pi/2-\theta$, the asymptotic
form of (\ref{eq:Log-mode}) for small $x$ becomes 
\begin{equation}
V_{p}\left(u,\theta\right)\sim-\frac{1}{3x^{2}}\log x+O\left(1\right),\label{eq:Asymptote_sph-wave}
\end{equation}
 which is same as the exact form of the Log mode of the AdS wave.
With the asymptotic behavior (\ref{eq:Asymptote_sph-wave}), the Log
mode associated with the spherical-AdS wave changes the asymptotically
AdS nature of the spacetime in the same way as the AdS wave.

Since the solutions we have found in this section are also solutions
of the linearized field equations as we show below, these metrics
constitute new explicit solutions for the Einstein and non-Einstein
(Log mode) excitations of the critical gravity besides the previously
studied AdS-wave solution \cite{Alishah,Gullu_Gurses}.

\section{Linearized Field Equations as Exact Field Equations}

Once one recognizes the fact that the curvature tensors, (\ref{eq:Riemann-KSK})
and (\ref{ein}), and the two tensors appearing in the field equations,
(\ref{eq:Box_Ricci_form1}-\ref{eq:Riem2}), are linear in the metric
function $V$ for the Kerr-Schild-Kundt (KSK) class of metrics defined
as 
\begin{equation}
g_{\mu\nu}=\bar{g}_{\mu\nu}+2V\lambda_{\mu}\lambda_{\nu},\qquad\lambda^{\mu}\partial_{\mu}V=0,\qquad\nabla_{\mu}\lambda_{\nu}=\lambda_{(\mu}\xi_{\nu)},\quad\lambda_{\mu}\xi^{\mu}=0,\label{eq:KSK-metric}
\end{equation}
 one realizes that the exact field equations of the quadratic curvature
gravity reduce to the linearized field equations in the metric perturbation
$h_{\mu\nu}\equiv g_{\mu\nu}-\bar{g}_{\mu\nu}=2V\lambda_{\mu}\lambda_{\nu}$
for the KSK class (\ref{eq:KSK-metric}). Even though this is straight
forward to see, let us analyze this observation in a little more detail
for the sake of completeness. First of all, for a generic metric perturbation
$h_{\mu\nu}$, the linearized field equations corresponding to the
field equations of the quadratic curvature gravity (\ref{fieldequations})
has the form \cite{DeserTekin,Deser_Tekin,Gullu_Tekin} 
\begin{equation}
c\,\mathcal{G}_{\mu\nu}^{L}+\left(2\alpha+\beta\right)\left(\bar{g}_{\mu\nu}\bar{\square}-\bar{\nabla}_{\mu}\bar{\nabla}_{\nu}+\frac{2\Lambda}{D-2}\bar{g}_{\mu\nu}\right)R^{L}+\beta\left(\bar{\square}\mathcal{G}_{\mu\nu}^{L}-\frac{2\Lambda}{D-1}\bar{g}_{\mu\nu}R^{L}\right)=0,\label{eq:Linearized_eom}
\end{equation}
 where the parameter $c$ is defined in (\ref{eq:c}), and $\mathcal{G}_{\mu\nu}^{L}$,
$R_{L}$ represent the linearized cosmological Einstein tensor and
the linearized scalar curvature, respectively, which have the forms
\begin{equation}
\mathcal{G}_{\mu\nu}^{L}=R_{\mu\nu}^{L}-\frac{1}{2}\bar{g}_{\mu\nu}R^{L}-\frac{2\Lambda}{D-2}h_{\mu\nu},
\end{equation}
 
\begin{equation}
R_{\mu\nu}^{L}=\frac{1}{2}\left(\bar{\nabla}^{\sigma}\bar{\nabla}_{\mu}h_{\nu\sigma}+\bar{\nabla}^{\sigma}\bar{\nabla}_{\nu}h_{\mu\sigma}-\bar{\square}h_{\mu\nu}-\bar{\nabla}_{\mu}\bar{\nabla}_{\nu}h\right),\qquad R^{L}=-\bar{\square}h+\bar{\nabla}^{\sigma}\bar{\nabla}^{\mu}h_{\sigma\mu}-\frac{2\Lambda}{D-2}h.
\end{equation}
 Here, $R_{\mu\nu}^{L}$ is the linearized Ricci tensor, and $\Lambda$
is the effective cosmological constant corresponding to the AdS background
and satisfies the field equation (\ref{quadratic}).

After describing the linearized field equations and the linearized
quantities for generic $h_{\mu\nu}$, let us focus on the KSK class
where $h_{\mu\nu}=2V\lambda_{\mu}\lambda_{\nu}$ and after this point
$h_{\mu\nu}$ represents the metric perturbation defined for the KSK
class. First thing to notice is that $h_{\mu\nu}$ satisfies $h=0$
and $\nabla_{\mu}h^{\mu\nu}=0$; therefore, the nontrivial part of
$h_{\mu\nu}$ is its transverse-traceless part which represents the
(massive and/or massless) spin-2 excitations. For tranverse-traceless
$h_{\mu\nu}$, the linearized field equations take the form 
\begin{align}
\left(\beta\bar{\square}+c\right)\mathcal{G}_{\mu\nu}^{L} & =0,\label{eq:Critic_short_eqn}
\end{align}
 where 
\begin{equation}
\mathcal{G}_{\mu\nu}^{L}=R_{\mu\nu}^{L}-\frac{2\Lambda}{D-2}h_{\mu\nu}=R_{\mu\nu}^{L}+k^{2}\left(D-1\right)h_{\mu\nu}.
\end{equation}
 Now, let us compare (\ref{eq:Critic_short_eqn}) with the quadratic
curvature gravity field equation for the KSK class (\ref{eq:Lambda_eqn_from_lin_eom}).
From (\ref{ein}), one can find the linearized Ricci tensor for KSK
class as 
\begin{equation}
R_{\mu\nu}^{L}=-\rho\lambda_{\mu}\lambda_{\nu}-k^{2}\left(D-1\right)h_{\mu\nu},
\end{equation}
 therefore, $\mathcal{G}_{\mu\nu}^{L}$ is just $\mathcal{G}_{\mu\nu}^{L}=-\rho\lambda_{\mu}\lambda_{\nu}$.
As a result, the field equations of the exact theory and the linearized
field equations are equivalent for the KSK class of metrics which
includes the AdS wave \cite{Gullu_Gurses} and the spherical-AdS wave
metrics presented above. Note that not all solutions of (\ref{eq:Critic_short_eqn})
taken as a linear equation of \emph{generic} perturbation $h_{\mu\nu}$
solve the full nonlinear theory. Such linear solutions were studied
in \cite{Bergshoeff-Log,Lu-LinModes}.

\section{Further Results and Conclusions}

We have defined a new subclass of metrics in Kerr-Schild-Kundt class
for which the null vector $\lambda^{\mu}$ has a symmetric covariant
derivative, namely $\nabla_{\mu}\lambda_{\nu}=\lambda_{(\mu}\xi_{\nu)}$
(note that $\lambda^{\mu}$ is not a recurrent vector; therefore,
our subclass does not have the special holonomy group ${\rm Sim}\left(n-2\right)$
discussed in \cite{Hervik}). Up to now two explicit metrics in this
class as solutions to quadratic gravity theories has been shown to
exist. One of them is the previously found AdS-wave metric \cite{Gullu_Gurses},
and the other one which we called spherical-AdS wave was presented
above. The latter solution is a generalization of the $D=3$ solution
of new massive gravity given in \cite{Aliev-PRL,Aliev-PLB}. Just
like the AdS wave, the spherical-AdS wave has Log modes which do not
asymptote to the AdS space \cite{Alishah,Gullu_Gurses}. As of now,
it is not clear if these two metrics exhaust the class of Kerr-Schild-Kundt
metrics having a null vector with a symmetric-covariant derivative
or there are some other.

In this work, even though we have concentrated in the quadratic gravity
theories both for the sake simplicity and for recent activity in quadratic
gravity theories, the class of metrics that we have studied has rather
remarkable properties which make them potential solutions to a large
class of theories that are built on arbitrary contractions of the
Riemann tensor whose Lagrangian is given as $f\left(g^{\mu\nu},R_{\mu\nu\rho\sigma}\right)$
along the lines of \cite{gurses-sisman-tekin-2011}. Leaving the details
for another work \cite{Gurses_f(Riem)}, let us summarize the curvature
properties of Kerr-Schild-Kundt class having a null vector with a
symmetric-covariant derivative: 
\begin{enumerate}
\item These metrics describe spacetimes with constant scalar invariants
built form the contractions of the Riemann tensor, but not its covariant
derivative, denoted as $\text{CSI}_{0}$ \cite{coley-hervik-pelavas-2006},
for example $R=-D\left(D-1\right)k^{2}$, $R_{\nu}^{\mu}R_{\mu}^{\nu}=D\left(D-1\right)^{2}k^{4}$,
$R_{\mu\alpha\beta\gamma}R^{\mu\alpha\beta\gamma}=2D(D-1)k^{4}$. 
\item All symmetric second rank tensors built from the contractions of the
Riemann tensor are linear in $\lambda_{\mu}\lambda_{\nu}$ for example
see (\ref{eq:Ricci2}-\ref{eq:Riem2}). This property implies property
1 above. This property is also sufficient to show that this class
of metrics also solve the Lovelock theory \cite{Gurses_f(Riem)}. 
\item Related to property 2, these metrics linearize the field equations.
For example, 
\begin{equation}
\square R_{\mu\nu}=\bar{\square}R_{\mu\nu}=-\lambda_{\mu}\lambda_{\nu}\left[\bar{\square}\rho+2\xi_{\mu}\partial^{\mu}\rho+\frac{1}{2}\rho\xi_{\mu}\xi^{\mu}-2\rho k^{2}\left(D-2\right)\right].
\end{equation}

\end{enumerate}
We expect that similar properties hold for symmetric two-tensors built
from the covariant derivatives of the Riemann tensor, namely $\left[\left(\nabla_{\gamma}^{\left(m\right)}R_{\mu\nu\rho\sigma}\right)^{n}\right]_{\alpha\beta}=a\left(k^{2}\right)g_{\alpha\beta}+b\left(\rho\right)\lambda_{\alpha}\lambda_{\beta}$,
which is consistent with the boost weight decomposition of the Riemann
tensor and its derivatives \cite{Hervik-BoostWeight} %
\footnote{We thank S.~Hervik for the discussion on this point.%
} This would lead to the result that these metrics could solve all
geometric theories.

\section{\label{ackno} Acknowledgments}

M.~G. is partially supported by the Scientific and Technological
Research Council of Turkey (TÜB\.{I}TAK). The work of T.~Ç.~\c{S}.
and B.~T. is supported by the TÜB\.{I}TAK Grant No.~110T339. We
would thank Sigbjørn Hervik and Tomáš Málek for their useful comments.
We thank a very conscientious referee whose useful remarks improved
the manuscript.

\appendix

\section{Definition of $\xi_{\nu}$ \label{sec:Definition-of-ksi}}

Let us discuss the symmetric-covariant derivative of the vector $\lambda^{\mu}$,
$\nabla_{\mu}\lambda_{\nu}=\lambda_{(\mu}\xi_{\nu)}$. Here, $\lambda_{\mu}\xi^{\mu}=0$
should hold in order to have $\lambda^{\mu}$ as a null geodesic.
Besides, note that $\nabla_{\mu}\lambda_{\nu}=\bar{\nabla}_{\mu}\lambda_{\nu}$
(see App.~\ref{sec:Curvature-Tensors-of-KS}). One can take the AdS
background metric in the canonical form as 
\begin{equation}
d\bar{s}^{2}=\frac{1}{k^{2}z^{2}}\left[-dt^{2}+\sum_{m=1}^{D-2}\left(dx^{m}\right)^{2}+dz^{2}\right],\label{eq:AdS_can}
\end{equation}
 where $z>0$ and $z\rightarrow0$ represents the AdS boundary. The
Christoffel connection of (\ref{eq:AdS_can}), which is in the form
$\bar{g}_{\mu\nu}=\omega^{-2}\eta_{\mu\nu}$ where $\omega\left(z\right)=kz$,
can be calculated with the usual conformal transformations as 
\begin{align}
\bar{\Gamma}_{\alpha\beta}^{\mu} & =\frac{1}{z}\eta_{\alpha\beta}\delta_{z}^{\mu}-\frac{1}{z}\left(\delta_{\alpha}^{\mu}\delta_{\beta}^{z}+\delta_{\beta}^{\mu}\delta_{\alpha}^{z}\right).
\end{align}
 With this result, $\bar{\nabla}_{\mu}\lambda_{\nu}$ becomes 
\begin{equation}
\bar{\nabla}_{\mu}\lambda_{\nu}=\partial_{\mu}\lambda_{\nu}-\frac{1}{z}\eta_{\mu\nu}\lambda_{z}+\frac{1}{z}\left(\lambda_{\mu}\delta_{\nu}^{z}+\lambda_{\nu}\delta_{\mu}^{z}\right).
\end{equation}
 Note that the last term in the parenthesis is already in the form
where $\lambda_{(\mu}\xi_{\nu)}$. Therefore, the first two terms
should take a form 
\begin{equation}
\partial_{\mu}\lambda_{\nu}-\frac{1}{z}\eta_{\mu\nu}\lambda_{z}=a\lambda_{\mu}\lambda_{\nu}+\lambda_{\mu}\zeta_{\nu}+\lambda_{\nu}\zeta_{\mu}.
\end{equation}

Now, let us define $\xi_{\mu}$ for the AdS-wave and the spherical-AdS
wave metrics. For AdS-wave metric, $\lambda_{\mu}$ has the form 
\begin{equation}
\lambda_{\mu}dx^{\mu}=\frac{1}{\sqrt{2}}\left(dt+dx\right),
\end{equation}
 in the canonical coordinates of AdS, and one has 
\begin{equation}
\bar{\nabla}_{\mu}\lambda_{\nu}=\frac{1}{z}\left(\lambda_{\mu}\delta_{\nu}^{z}+\lambda_{\nu}\delta_{\mu}^{z}\right)\Rightarrow\xi_{\mu}=\frac{2}{z}\delta_{\mu}^{z}.
\end{equation}
 For the spherical-AdS wave, one has 
\begin{equation}
\lambda_{\mu}dx^{\mu}=dt+\sum_{m=1}^{D-2}\frac{x^{m}}{r}dx^{m}+\frac{z}{r}dz,\qquad r^{2}=\sum_{m=1}^{D-2}\left(x^{m}\right)^{2}+z^{2},
\end{equation}
 and $\bar{\nabla}_{\mu}\lambda_{\nu}$ becomes 
\begin{equation}
\bar{\nabla}_{\mu}\lambda_{\nu}=-\frac{1}{r}\lambda_{\mu}\lambda_{\nu}+\frac{1}{r}\delta_{\mu}^{t}\lambda_{\nu}+\frac{1}{r}\delta_{\nu}^{t}\lambda_{\mu}+\frac{1}{z}\left(\lambda_{\mu}\delta_{\nu}^{z}+\lambda_{\nu}\delta_{\mu}^{z}\right)
\end{equation}
 therefore, 
\begin{equation}
\xi_{\mu}=-\frac{1}{r}\lambda_{\mu}+\frac{2}{r}\delta_{\mu}^{t}+\frac{2}{z}\delta_{\mu}^{z}.
\end{equation}

\section{Curvature Tensors of the Kerr-Schild Metric\label{sec:Curvature-Tensors-of-KS}}

In this section, we obtain the forms of the Riemann and Ricci tensors,
and the scalar curvature for the Kerr-Schild metric 
\begin{equation}
g_{\mu\nu}=\bar{g}_{\mu\nu}+2V\lambda_{\mu}\lambda_{\nu},\label{eq:KS-metric}
\end{equation}
 where $\bar{g}_{\mu\nu}$ is the metric of the AdS spacetime, the
vector $\lambda^{\mu}$ is null and geodesic for both $g_{\mu\nu}$
and $\bar{g}_{\mu\nu}$; 
\begin{equation}
\lambda_{\mu}\lambda^{\mu}=g_{\mu\nu}\lambda^{\mu}\lambda^{\nu}=\bar{g}_{\mu\nu}\lambda^{\mu}\lambda^{\nu}=0,\label{eq:KS-null}
\end{equation}
\begin{equation}
\lambda^{\mu}\nabla_{\mu}\lambda_{\rho}=\lambda^{\mu}\bar{\nabla}_{\mu}\lambda_{\rho}=0,\label{eq:KS-geo}
\end{equation}
 and, finally, $V$ is a function of spacetime which is assumed to
satisfy $\lambda^{\mu}\partial_{\mu}V=0$.%
\footnote{The exposition until Appendix~B.1 is rather standard. Here, we provide
self-contained presentation on curvature tensors of the Kerr-Schild
metric (\ref{eq:KS-metric}) satisfying $\lambda^{\mu}\partial_{\mu}V=0$
in addition to the generally assumed properties (\ref{eq:KS-null})
and (\ref{eq:KS-geo}). See \cite{Xanthopoulos,Chandra} for KS metrics
having the property (\ref{eq:KS-null}) with a flat background and
\cite{Anabalon} for KS satisfying (\ref{eq:KS-null}) and (\ref{eq:KS-geo})
for generic backgrounds and generic $V$.%
} The Christoffel connection of $g_{\mu\nu}$ has the form 
\begin{equation}
\Gamma_{\alpha\beta}^{\mu}=\bar{\Gamma}_{\alpha\beta}^{\mu}+\Omega_{\phantom{\mu}\alpha\beta}^{\mu},\label{eq:Christoffel_KS}
\end{equation}
 where $\bar{\Gamma}_{\alpha\beta}^{\mu}$ is the Christoffel connection
of the background metric $\bar{g}_{\mu\nu}$, and the terms linear
in $V$ collected in $\Omega_{\phantom{\mu}\alpha\beta}^{\mu}$ which
can be written as 
\begin{equation}
\Omega_{\phantom{\mu}\alpha\beta}^{\mu}=\bar{\nabla}_{\alpha}\left(V\lambda^{\mu}\lambda_{\beta}\right)+\bar{\nabla}_{\beta}\left(V\lambda^{\mu}\lambda_{\alpha}\right)-\bar{\nabla}^{\mu}\left(V\lambda_{\alpha}\lambda_{\beta}\right).\label{eq:Lin_Chris}
\end{equation}
 One can easily show that $\Omega_{\phantom{\mu}\alpha\beta}^{\mu}$
satisfies the properties 
\begin{equation}
\Omega_{\phantom{\mu}\mu\beta}^{\mu}=0,\qquad\lambda_{\mu}\Omega_{\phantom{\mu}\alpha\beta}^{\mu}=0,\qquad\lambda^{\alpha}\Omega_{\phantom{\mu}\alpha\beta}^{\mu}=0,\label{eq:Christoffel_properties}
\end{equation}
 which have the important implication that the covariant derivative
of $\lambda^{\mu}$ reduces to the covariant derivative with respect
to the background AdS metric, namely 
\begin{equation}
\nabla_{\mu}\lambda_{\rho}=\bar{\nabla}_{\mu}\lambda_{\rho}.
\end{equation}
 With (\ref{eq:Christoffel_KS}), the Riemann tensor has the form
\begin{equation}
R_{\phantom{\mu}\alpha\nu\beta}^{\mu}=\bar{R}_{\phantom{\mu}\alpha\nu\beta}^{\mu}+\bar{\nabla}_{\nu}\Omega_{\phantom{\mu}\alpha\beta}^{\mu}-\bar{\nabla}_{\beta}\Omega_{\phantom{\mu}\alpha\nu}^{\mu}+\Omega_{\phantom{\mu}\nu\sigma}^{\mu}\Omega_{\phantom{\sigma}\beta\alpha}^{\sigma}-\Omega_{\phantom{\mu}\beta\sigma}^{\mu}\Omega_{\phantom{\sigma}\nu\alpha}^{\sigma},\label{eq:Riemann_KS}
\end{equation}
 where $\bar{R}_{\phantom{\mu}\alpha\nu\beta}^{\mu}$ is the Riemann
tensor of the AdS spacetime having the form 
\begin{equation}
\bar{R}_{\phantom{\mu}\alpha\nu\beta}^{\mu}=-k^{2}\left(\delta_{\nu}^{\mu}\bar{g}_{\alpha\beta}-\delta_{\beta}^{\mu}\bar{g}_{\alpha\nu}\right).\label{eq:AdS_Riem}
\end{equation}
 Contraction of the Riemann tensor with two $\lambda^{\mu}$ vectors
has the simple form 
\begin{equation}
\lambda_{\mu}\lambda^{\nu}R_{\phantom{\mu}\alpha\nu\beta}^{\mu}=\lambda_{\mu}\lambda^{\nu}\bar{R}_{\phantom{\mu}\alpha\nu\beta}^{\mu}=k^{2}\lambda_{\alpha}\lambda_{\beta},\qquad\lambda^{\alpha}\lambda^{\nu}R_{\phantom{\mu}\alpha\nu\beta}^{\mu}=\lambda^{\alpha}\lambda^{\nu}\bar{R}_{\phantom{\mu}\alpha\nu\beta}^{\mu}=-k^{2}\lambda^{\mu}\lambda_{\beta}.
\end{equation}
 Using (\ref{eq:Christoffel_properties}), one can obtain the Ricci
tensor from (\ref{eq:Riemann_KS}) as 
\begin{equation}
R_{\alpha\beta}=\bar{R}_{\alpha\beta}+\bar{\nabla}_{\mu}\Omega_{\phantom{\mu}\alpha\beta}^{\mu}-\Omega_{\phantom{\mu}\beta\sigma}^{\mu}\Omega_{\phantom{\sigma}\mu\alpha}^{\sigma},\label{eq:Ricci_down_KS}
\end{equation}
 where the Ricci tensor of the AdS spacetime is $\bar{R}_{\alpha\beta}=-k^{2}\left(D-1\right)\bar{g}_{\alpha\beta}$.
The last term can be written in the form 
\begin{align}
\Omega_{\phantom{\mu}\beta\sigma}^{\mu}\Omega_{\phantom{\sigma}\mu\alpha}^{\sigma}= & -4V^{2}\lambda_{\alpha}\lambda_{\beta}\left(\bar{\nabla}_{[\mu}\lambda_{\sigma]}\right)\bar{\nabla}^{\mu}\lambda^{\sigma},
\end{align}
 therefore the Ricci tensor with down indices is quadratic in $V$.
However, it is well-known that the Ricci tensor with up-down indices,
$R_{\beta}^{\rho}=g^{\rho\alpha}R_{\alpha\beta}$, is linear in $V$
for a metric in the Kerr-Schild form \cite{DereliGurses}; 
\begin{equation}
R_{\beta}^{\rho}=\bar{R}_{\beta}^{\rho}-2V\lambda^{\rho}\lambda^{\alpha}\bar{R}_{\alpha\beta}+\bar{g}^{\rho\alpha}\bar{\nabla}_{\mu}\Omega_{\phantom{\mu}\alpha\beta}^{\mu}.\label{eq:Ricci_up-down_KS}
\end{equation}
 Finally, the scalar curvature is a constant having a value which
is equal to the background one; 
\begin{equation}
R=\bar{R}=-D\left(D-1\right)k^{2}.\label{eq:R_KS}
\end{equation}

\subsection{Curvature tensors of the Kerr-Schild-Kundt class\label{sub:Curvature-tensors-of-KSK}}

Up to now, we consider the Kerr-Schild metrics for which $\lambda^{\mu}$
is a null geodesic as usual. On the other hand, the AdS-wave and spherical-AdS-wave
metrics belong to the class of Kerr-Schild-Kundt (KSK) metrics for
which the vector $\lambda^{\mu}$ satisfies the property 
\begin{equation}
\nabla_{\mu}\lambda_{\nu}=\lambda_{(\mu}\xi_{\nu)},\qquad\xi^{\mu}\lambda_{\mu}=0.\label{eq:Class_defn}
\end{equation}
 Note that due to $\xi^{\mu}\lambda_{\mu}=0$, one has $\xi^{\mu}=g^{\mu\nu}\xi_{\nu}=\bar{g}^{\mu\nu}\xi_{\nu}$.
The non-expanding, $\nabla_{\mu}\lambda^{\mu}=0$, shear-free, $\nabla^{\mu}\lambda^{\nu}\nabla_{(\mu}\lambda_{\nu)}=0$,
and non-twisting, $\nabla^{\mu}\lambda^{\nu}\nabla_{[\mu}\lambda_{\nu]}=0$,
nature of the vector $\lambda^{\mu}$ simply follows from (\ref{eq:Class_defn})
which means the Kerr-Schild metric is a member of Kundt class by definition.
Immediate implications of (\ref{eq:Class_defn}) are 
\begin{equation}
\xi^{\mu}\nabla_{\mu}\lambda_{\nu}=\xi^{\mu}\nabla_{\nu}\lambda_{\mu}=\frac{1}{2}\lambda_{\nu}\xi^{\mu}\xi_{\mu},
\end{equation}
 and 
\begin{equation}
\bar{\nabla}_{\nu}\left(\xi^{\mu}\lambda_{\mu}\right)=0\Rightarrow\lambda^{\mu}\bar{\nabla}_{\nu}\xi_{\mu}=-\xi^{\mu}\bar{\nabla}_{\nu}\lambda_{\mu}.
\end{equation}
 Using the Ricci identity in the form $\left[\bar{\nabla}_{\mu},\bar{\nabla}_{\nu}\right]\lambda^{\mu}=\bar{R}_{\nu\sigma}\lambda^{\sigma}$
together with $\bar{\nabla}_{\mu}\lambda^{\mu}=0$, one can obtain
\begin{equation}
\bar{\square}\lambda_{\nu}=-k^{2}\left(D-1\right)\lambda_{\nu},\label{eq:Box_lambda-KSK}
\end{equation}
 and explicitly calculating the left-hand side yields the relation
\begin{equation}
\lambda^{\mu}\bar{\nabla}_{\mu}\xi_{\nu}=-\lambda_{\nu}\left[\bar{\nabla}_{\mu}\xi^{\mu}+\frac{1}{2}\xi^{\mu}\xi_{\mu}+2k^{2}\left(D-1\right)\right]\label{eq:D_bar_ksi_in_lambda_dir}
\end{equation}
 that is used in the calculations below.

In order to study the curvature tensors, first one should find the
$\Omega_{\phantom{\mu}\alpha\beta}^{\mu}$ part of the Christoffel
connection which is linear in $V$, and it becomes 
\begin{equation}
\Omega_{\phantom{\mu}\alpha\beta}^{\mu}=-\lambda_{\alpha}\lambda_{\beta}\partial^{\mu}V+2\lambda^{\mu}\lambda_{(\alpha}\partial_{\beta)}V+2V\lambda^{\mu}\lambda_{(\alpha}\xi_{\beta)}.\label{eq:Lin_Chris-KSK}
\end{equation}
 Note that contraction of the vector $\xi^{\mu}$ with $\Omega_{\phantom{\mu}\alpha\beta}^{\mu}$
yields 
\begin{equation}
\xi_{\mu}\Omega_{\phantom{\mu}\alpha\beta}^{\mu}=-\lambda_{\alpha}\lambda_{\beta}\xi_{\mu}\partial^{\mu}V,\qquad\xi^{\alpha}\Omega_{\phantom{\mu}\alpha\beta}^{\mu}=\lambda^{\mu}\lambda_{\beta}\left(\xi^{\alpha}\partial_{\alpha}V+V\xi^{\alpha}\xi_{\alpha}\right),\label{eq:ksi_contracts_Lin-Chris}
\end{equation}
 so $\nabla_{\mu}\xi_{\rho}\ne\bar{\nabla}_{\mu}\xi_{\rho}.$ Now,
using (\ref{eq:Lin_Chris-KSK}), we can calculate $\bar{\nabla}_{\nu}\Omega_{\phantom{\mu}\alpha\beta}^{\mu}$
and $\Omega_{\phantom{\mu}\nu\sigma}^{\mu}\Omega_{\phantom{\sigma}\beta\alpha}^{\sigma}$
for KSK class. First, $\bar{\nabla}_{\nu}\Omega_{\phantom{\mu}\alpha\beta}^{\mu}$
can be obtained as 
\begin{align}
\bar{\nabla}_{\nu}\Omega_{\phantom{\mu}\alpha\beta}^{\mu}= & -\lambda_{\alpha}\lambda_{\beta}\left(\bar{\nabla}_{\nu}\partial^{\mu}V+\xi_{\nu}\partial^{\mu}V\right)-\lambda_{\nu}\lambda_{(\alpha}\xi_{\beta)}\partial^{\mu}V\nonumber \\
 & +2\lambda^{\mu}\lambda_{(\alpha}\bar{\nabla}_{\beta)}\partial_{\nu}V+2\lambda^{\mu}\lambda_{(\alpha}\xi_{\beta)}\partial_{\nu}V+2V\lambda^{\mu}\lambda_{(\alpha|}\bar{\nabla}_{\nu}\xi_{|\beta)}\nonumber \\
 & +\left(2\lambda^{\mu}\xi_{\nu}+\xi^{\mu}\lambda_{\nu}\right)\left(V\lambda_{(\alpha}\xi_{\beta)}+\lambda_{(\alpha}\partial_{\beta)}V\right)\nonumber \\
 & +\lambda^{\mu}\lambda_{\nu}\left(V\xi_{\alpha}\xi_{\beta}+\xi_{(\alpha}\partial_{\beta)}V\right).\label{eq:D_bar_Lin-Chris}
\end{align}
 Then, the linear in $V$ terms in the Riemann tensor becomes 
\begin{align}
\bar{\nabla}_{\nu}\Omega_{\phantom{\mu}\alpha\beta}^{\mu}-\bar{\nabla}_{\beta}\Omega_{\phantom{\mu}\alpha\nu}^{\mu}= & 2\lambda_{\alpha}\lambda_{[\nu}\bar{\nabla}_{\beta]}\partial^{\mu}V-2\lambda^{\mu}\lambda_{[\nu}\bar{\nabla}_{\beta]}\partial_{\alpha}V\nonumber \\
 & +\lambda_{[\nu}\xi_{\beta]}\left(\lambda_{\alpha}\partial^{\mu}V-\lambda^{\mu}\partial_{\alpha}V+\lambda_{\alpha}\xi^{\mu}V\right)\nonumber \\
 & +\left(\lambda_{\alpha}\xi^{\mu}-\lambda^{\mu}\xi_{\alpha}\right)\lambda_{[\nu}\partial_{\beta]}V\nonumber \\
 & +2V\lambda^{\mu}\left(\lambda_{\alpha}\bar{\nabla}_{[\nu}\xi_{\beta]}-\lambda_{[\nu}\bar{\nabla}_{\beta]}\xi_{\alpha}\right).\label{eq:Riem_L-KSK}
\end{align}
 Secondly, the term $\Omega_{\phantom{\mu}\nu\sigma}^{\mu}\Omega_{\phantom{\sigma}\beta\alpha}^{\sigma}$
has the form 
\begin{equation}
\Omega_{\phantom{\mu}\nu\sigma}^{\mu}\Omega_{\phantom{\sigma}\beta\alpha}^{\sigma}=-\lambda^{\mu}\lambda_{\alpha}\lambda_{\beta}\lambda_{\nu}\left(\partial^{\sigma}V\right)\left(V\xi_{\sigma}+\partial_{\sigma}V\right).
\end{equation}
 Note that $\Omega_{\phantom{\mu}\nu\sigma}^{\mu}\Omega_{\phantom{\sigma}\beta\alpha}^{\sigma}$
is symmetric in $\nu$ and $\beta$ indices; therefore, the quadratic
in $V$ terms in the Riemann tensor cancel each other due to antisymmetry
in $\nu$ and $\beta$. Thus, the Riemann tensor for the KSK class
is linear in $V$ and has the form 
\begin{equation}
R_{\phantom{\mu}\alpha\nu\beta}^{\mu}=\bar{R}_{\phantom{\mu}\alpha\nu\beta}^{\mu}+\bar{\nabla}_{\nu}\Omega_{\phantom{\mu}\alpha\beta}^{\mu}-\bar{\nabla}_{\beta}\Omega_{\phantom{\mu}\alpha\nu}^{\mu},\label{eq:Riem-KSK}
\end{equation}
 where the last two terms are given in (\ref{eq:Riem_L-KSK}). Now,
let us discuss the contractions of the Riemann tensor with one $\lambda^{\mu}$
vector. By using (\ref{eq:Christoffel_properties}), (\ref{eq:Class_defn})
and (\ref{eq:ksi_contracts_Lin-Chris}), one can show that 
\begin{align}
\lambda_{\mu}R_{\phantom{\mu}\alpha\nu\beta}^{\mu}= & \lambda_{\mu}\bar{R}_{\phantom{\mu}\alpha\nu\beta}^{\mu},\qquad\lambda^{\alpha}R_{\phantom{\mu}\alpha\nu\beta}^{\mu}=\lambda^{\alpha}\bar{R}_{\phantom{\mu}\alpha\nu\beta}^{\mu},\qquad\lambda^{\nu}R_{\phantom{\mu}\alpha\nu\beta}^{\mu}=\lambda^{\nu}\bar{R}_{\phantom{\mu}\alpha\nu\beta}^{\mu}-2k^{2}V\lambda^{\mu}\lambda_{\alpha}\lambda_{\beta},\label{eq:lambda-Riem}
\end{align}
 where the last one is implied by either one of the previous two results.
After using (\ref{eq:AdS_Riem}), one can also have 
\begin{equation}
\lambda_{\mu}R_{\phantom{\mu}\alpha\nu\beta}^{\mu}=\frac{R}{D\left(D-1\right)}\left(\lambda_{\nu}g_{\alpha\beta}-\lambda_{\beta}g_{\alpha\nu}\right),
\end{equation}
 where the right-hand side can also be written in terms of background
quantities, and the other two contractions follow similarly. On the
other hand, one can calculate $\lambda^{\nu}R_{\phantom{\mu}\alpha\nu\beta}^{\mu}$
explicitly by using (\ref{eq:Christoffel_properties}), (\ref{eq:Class_defn}),
(\ref{eq:D_bar_Lin-Chris}), (\ref{eq:ksi_contracts_Lin-Chris}) and
(\ref{eq:D_bar_ksi_in_lambda_dir}) as 
\begin{equation}
\lambda^{\nu}R_{\phantom{\mu}\alpha\nu\beta}^{\mu}=\lambda^{\nu}\bar{R}_{\phantom{\mu}\alpha\nu\beta}^{\mu}-2V\lambda^{\mu}\lambda_{\alpha}\lambda_{\beta}\left(\bar{\nabla}_{\nu}\xi^{\nu}+\frac{1}{4}\xi_{\nu}\xi^{\nu}+2k^{2}\left(D-1\right)\right),
\end{equation}
 which together with (\ref{eq:lambda-Riem}) implies 
\begin{equation}
\bar{\nabla}_{\nu}\xi^{\nu}+\frac{1}{4}\xi_{\nu}\xi^{\nu}+k^{2}\left(2D-3\right)=0.\label{eq:div-ksi_ksi^2_reln}
\end{equation}
 This relation can be verified explicitly for the AdS wave and the
spherical-AdS wave cases.

In order to calculate the Ricci tensor, one needs to calculate $\bar{\nabla}_{\mu}\Omega_{\phantom{\mu}\alpha\beta}^{\mu}$.
One may follow two routes: directly computing it from (\ref{eq:D_bar_Lin-Chris})
by using (\ref{eq:D_bar_ksi_in_lambda_dir}) and (\ref{eq:div-ksi_ksi^2_reln})
or using the following result obtained by use of the Ricci identity;
\begin{equation}
\bar{\nabla}_{\mu}\bar{\nabla}_{\alpha}\left(V\lambda^{\mu}\lambda_{\beta}\right)=-k^{2}DV\lambda_{\alpha}\lambda_{\beta},
\end{equation}
 with the original form of the $\Omega_{\phantom{\mu}\alpha\beta}^{\mu}$
in (\ref{eq:Lin_Chris}). Then, one can obtain the Ricci tensor as
\begin{equation}
R_{\alpha\beta}=-k^{2}\left(D-1\right)g_{\alpha\beta}-\rho\lambda_{\alpha}\lambda_{\beta},\label{eq:Ricci-KSK_with_rho}
\end{equation}
 where 
\begin{equation}
\rho\equiv\bar{\square}V+2\xi_{\mu}\partial^{\mu}V+\frac{1}{2}V\xi_{\mu}\xi^{\mu}-2Vk^{2}\left(D-2\right),\label{eq:rho-KSK}
\end{equation}
 or 
\begin{equation}
R_{\alpha\beta}=-k^{2}\left(D-1\right)g_{\alpha\beta}-\left(\bar{\square}+2k^{2}\right)\left(V\lambda_{\alpha}\lambda_{\beta}\right).\label{eq:Ricci-KSK}
\end{equation}
 Two forms of the Ricci tensor imply 
\begin{equation}
\bar{\square}\left(V\lambda_{\alpha}\lambda_{\beta}\right)=\left(\rho-2Vk^{2}\right)\lambda_{\alpha}\lambda_{\beta}.
\end{equation}
 It is possible to verify this relation by explicitly calculating
the left-hand side by using (\ref{eq:Box_lambda-KSK}). Besides, one
can easily show that the scalar curvature is constant, since the linear
part of the Ricci tensor is in the form $R_{\alpha\beta}^{L}\sim\lambda_{\alpha}\lambda_{\beta}$.

Finally, let us show that the KSK metrics satisfy $C_{\mu\alpha\nu\beta}\lambda^{\beta}=0$
where the Weyl tensor is defined as 
\begin{equation}
C_{\mu\alpha\nu\beta}\equiv R_{\mu\alpha\nu\beta}-\frac{2}{D-2}\left(g_{\mu[\nu}R_{\beta]\alpha}-g_{\alpha[\nu}R_{\beta]\mu}\right)+\frac{2}{\left(D-1\right)\left(D-2\right)}Rg_{\mu[\nu}g_{\beta]\alpha}.
\end{equation}
 Using (\ref{eq:lambda-Riem}), $g_{\mu\nu}-\bar{g}_{\mu\nu}\sim\lambda_{\mu}\lambda_{\nu}$
and $R_{\mu\nu}-\bar{R}_{\mu\nu}\sim\lambda_{\mu}\lambda_{\nu}$,
it can be shown that $C_{\mu\alpha\nu\beta}\lambda^{\beta}$ reduces
to $\bar{C}_{\mu\alpha\nu\beta}\lambda^{\beta}$ where $\bar{C}_{\mu\alpha\nu\beta}=0$;
therefore, one has 
\begin{equation}
C_{\mu\alpha\nu\beta}\lambda^{\beta}=\bar{C}_{\mu\alpha\nu\beta}\lambda^{\beta}=0.
\end{equation}

\subsection{Two tensors in the field equations}

In order to find the field equations of the quadratic curvature gravity
for the KSK metrics (\ref{eq:Class_defn}), one needs to obtain the
form of the two tensors $R_{\mu}^{\rho}R_{\rho\nu}$, $R_{\mu\alpha\nu\beta}R^{\alpha\beta}$,
$R_{\mu\alpha\beta\gamma}R_{\nu}^{\phantom{\nu}\alpha\beta\gamma}$
and $\square R_{\mu\nu}$ for this class of metrics. By using (\ref{eq:Ricci-KSK_with_rho}),
the term $R_{\mu}^{\rho}R_{\rho\nu}$ can easily be calculated as
\begin{equation}
R_{\mu}^{\rho}R_{\rho\nu}=\left(D-1\right)^{2}k^{4}g_{\mu\nu}+2\left(D-1\right)k^{2}\rho\lambda_{\mu}\lambda_{\nu}.
\end{equation}
 The term $R_{\mu\alpha\nu\beta}R^{\alpha\beta}$ is also rather simple:
after using (\ref{eq:Ricci-KSK_with_rho}) and (\ref{eq:lambda-Riem}),
one has 
\begin{equation}
R_{\mu\alpha\nu\beta}R^{\alpha\beta}=\left(D-1\right)^{2}k^{4}g_{\mu\nu}+\left(D-2\right)k^{2}\rho\lambda_{\mu}\lambda_{\nu}.
\end{equation}
 Then, moving to $R_{\mu\alpha\beta\gamma}R_{\nu}^{\phantom{\nu}\alpha\beta\gamma}$
whose calculation is straightforward, but time consuming. It is better
to calculate $R_{\phantom{\mu}\sigma\nu\beta}^{\mu}R_{\gamma}^{\phantom{\gamma}\sigma\nu\beta}=R_{\phantom{\mu\sigma}\nu\beta}^{\mu\sigma}R_{\phantom{\nu\beta}\gamma\sigma}^{\nu\beta}$
which can be written as 
\begin{equation}
R_{\phantom{\mu\sigma}\nu\beta}^{\mu\sigma}R_{\phantom{\nu\beta}\gamma\sigma}^{\nu\beta}=R_{\phantom{\nu\beta}\gamma\sigma}^{\nu\beta}\bar{g}^{\sigma\alpha}R_{\phantom{\mu}\alpha\nu\beta}^{\mu}-2V\bar{R}_{\phantom{\nu\beta}\gamma\sigma}^{\nu\beta}\lambda^{\sigma}\lambda^{\alpha}R_{\phantom{\mu}\alpha\nu\beta}^{\mu},
\end{equation}
 where (\ref{eq:lambda-Riem}) is used and the first term explicitly
has the form 
\begin{align}
R_{\phantom{\nu\beta}\gamma\sigma}^{\nu\beta}\bar{g}^{\sigma\alpha}R_{\phantom{\mu}\alpha\nu\beta}^{\mu}=R_{\phantom{\nu\beta}\gamma\sigma}^{\nu\beta}\bar{g}^{\sigma\alpha}\biggl\{ & \bar{R}_{\phantom{\mu}\alpha\nu\beta}^{\mu}+2\lambda_{\alpha}\lambda_{[\nu}\bar{\nabla}_{\beta]}\partial^{\mu}V-2\lambda^{\mu}\lambda_{[\nu}\bar{\nabla}_{\beta]}\partial_{\alpha}V\nonumber \\
 & +\lambda_{[\nu}\xi_{\beta]}\left(\lambda_{\alpha}\partial^{\mu}V-\lambda^{\mu}\partial_{\alpha}V+\xi^{\mu}\lambda_{\alpha}V\right)\nonumber \\
 & +\left(\lambda_{\alpha}\xi^{\mu}-\lambda^{\mu}\xi_{\alpha}\right)\lambda_{[\nu}\partial_{\beta]}V\nonumber \\
 & +2V\lambda^{\mu}\left(\lambda_{\alpha}\bar{\nabla}_{[\nu}\xi_{\beta]}-\lambda_{[\nu}\bar{\nabla}_{\beta]}\xi_{\alpha}\right)\biggr\}.
\end{align}
 Since the terms in $R_{\phantom{\mu}\alpha\nu\beta}^{\mu}$ which
are linear in $V$ involve either $\lambda_{\alpha}$ or $\lambda_{\nu}$
or $\lambda_{\beta}$, using again (\ref{eq:lambda-Riem}) yields
\begin{equation}
R_{\phantom{\nu\beta}\gamma\sigma}^{\nu\beta}\bar{g}^{\sigma\alpha}R_{\phantom{\mu}\alpha\nu\beta}^{\mu}=\bar{R}_{\phantom{\nu\beta}\gamma\sigma}^{\nu\beta}\bar{g}^{\sigma\alpha}R_{\phantom{\mu}\alpha\nu\beta}^{\mu}+\left(R_{\phantom{\nu\beta}\gamma\sigma}^{\nu\beta}\right)_{L}\bar{R}_{\phantom{\mu\sigma}\nu\beta}^{\mu\sigma},
\end{equation}
 where $\left(R_{\phantom{\nu\beta}\gamma\sigma}^{\nu\beta}\right)_{L}\equiv R_{\phantom{\nu\beta}\gamma\sigma}^{\nu\beta}-\bar{R}_{\phantom{\nu\beta}\gamma\sigma}^{\nu\beta}$.
With this result and (\ref{eq:AdS_Riem}), $R_{\phantom{\mu\sigma}\nu\beta}^{\mu\sigma}R_{\phantom{\nu\beta}\gamma\sigma}^{\nu\beta}$
becomes 
\begin{align}
R_{\phantom{\mu\sigma}\nu\beta}^{\mu\sigma}R_{\phantom{\nu\beta}\gamma\sigma}^{\nu\beta}= & \bar{R}_{\phantom{\nu\beta}\gamma\sigma}^{\nu\beta}R_{\phantom{\mu\sigma}\nu\beta}^{\mu\sigma}+\left(R_{\phantom{\nu\beta}\gamma\sigma}^{\nu\beta}\right)_{L}\bar{R}_{\phantom{\mu\sigma}\nu\beta}^{\mu\sigma}=-2k^{2}\left[R_{\gamma}^{\mu}+2k^{2}\left(R_{\gamma}^{\mu}\right)_{L}\right],
\end{align}
 where $\left(R_{\gamma}^{\mu}\right)_{L}=-\rho\lambda^{\mu}\lambda_{\gamma}$
from (\ref{eq:Ricci-KSK_with_rho}). As a result, one obtains 
\begin{equation}
R_{\mu\alpha\beta\gamma}R_{\nu}^{\phantom{\nu}\alpha\beta\gamma}=2(D-1)k^{4}g_{\mu\nu}+4k^{2}\rho\lambda_{\mu}\lambda_{\nu}.
\end{equation}

Finally, let us study the term $\square R_{\mu\nu}$, and from (\ref{eq:Ricci-KSK_with_rho})
it immediately becomes $\square R_{\mu\nu}=-\square\left(\rho\lambda_{\alpha}\lambda_{\beta}\right)$.
Then, since $\nabla_{\mu}\lambda_{\rho}=\bar{\nabla}_{\mu}\lambda_{\rho}$,
$\square\lambda^{\mu}=\bar{\square}\lambda^{\mu}$ and $\square\rho=\bar{\square}\rho$,
one has 
\begin{equation}
\square R_{\mu\nu}=\bar{\square}R_{\mu\nu}=-\bar{\square}\left(\rho\lambda_{\mu}\lambda_{\nu}\right).
\end{equation}
 In App.~\ref{sub:Curvature-tensors-of-KSK}, we have discussed the
explicit calculation of $\bar{\square}\left(V\lambda_{\mu}\lambda_{\nu}\right)$
which becomes 
\begin{equation}
\bar{\square}\left(V\lambda_{\mu}\lambda_{\nu}\right)=\lambda_{\mu}\lambda_{\nu}\left(\bar{\square}V+2\xi_{\sigma}\partial^{\sigma}V+\frac{1}{2}V\xi_{\sigma}\xi^{\sigma}-2Vk^{2}\left(D-1\right)\right),
\end{equation}
 and in deriving this relation $\lambda^{\mu}\partial_{\mu}V=0$ is
used. One can show that $\lambda^{\mu}\partial_{\mu}\rho=0$ (note
that $\lambda^{\sigma}\bar{\nabla}_{\sigma}\xi_{\mu}\sim\lambda_{\mu}$),
then the same relation also holds for $\rho$. Hence, one has 
\begin{equation}
\square R_{\mu\nu}=-\lambda_{\mu}\lambda_{\nu}\left(\bar{\square}\rho+2\xi_{\sigma}\partial^{\sigma}\rho+\frac{1}{2}\rho\xi_{\sigma}\xi^{\sigma}-2\rho k^{2}\left(D-1\right)\right).
\end{equation}

\section{Spherical-AdS Wave Computations}

Let us have the AdS metric in the coordinates 
\begin{equation}
d\bar{s}^{2}=\frac{4dudv}{k^{2}\cos^{2}\theta\left(u+v\right)^{2}}+\frac{1}{k^{2}\cos^{2}\theta}d\Omega_{D-2}^{2}.
\end{equation}
 Then, some components of the Christoffel connection for this metric
are 
\begin{align}
\bar{\Gamma}_{uu}^{u} & =-\frac{2}{u+v},\qquad\bar{\Gamma}_{uv}^{u}=0,\qquad\bar{\Gamma}_{u\theta}^{u}=\tan\theta,\qquad\bar{\Gamma}_{\theta\theta}^{u}=0\nonumber \\
\bar{\Gamma}_{uv}^{v} & =0,\qquad\bar{\Gamma}_{\theta\theta}^{v}=0,\qquad\bar{\Gamma}_{uv}^{\theta}=-\frac{2\tan\theta}{\left(u+v\right)^{2}},\qquad\bar{\Gamma}_{\theta\theta}^{\theta}=\tan\theta,\label{eq:Chris_AdS_sph}\\
\bar{\Gamma}_{\Omega_{i}\Omega_{i}}^{\theta} & =-k^{2}\bar{g}_{\Omega_{i}\Omega_{i}}\cot\theta,\qquad\bar{\Gamma}_{\Omega_{i}\Omega_{i}}^{u}=0,\qquad\bar{\Gamma}_{\Omega_{i}\Omega_{i}}^{v}=0.\nonumber 
\end{align}
 where $\Omega_{i}$ denotes the angular coordinates of $d\Omega_{D-2}^{2}$
other than $\theta$. Now, let us first discuss the form of $\bar{\nabla}_{\mu}\lambda_{\nu}$;
\begin{equation}
\bar{\nabla}_{\mu}\lambda_{\nu}=-\bar{\Gamma}_{\mu\nu}^{u}=-\lambda_{\mu}\lambda_{\nu}\bar{\Gamma}_{uu}^{u}-\lambda_{\mu}\bar{\Gamma}_{u\theta}^{u}\delta_{\nu}^{\theta}-\delta_{\mu}^{\theta}\bar{\Gamma}_{\theta u}^{u}\lambda_{\nu}
\end{equation}
 and one has 
\begin{equation}
\xi_{\nu}\equiv-\bar{\Gamma}_{uu}^{u}\lambda_{\nu}-2\bar{\Gamma}_{u\theta}^{u}\delta_{\nu}^{\theta}.
\end{equation}
 Finally, one can calculate $\rho$ as 
\begin{equation}
\rho=\bar{\square}V-4k^{2}\sin\theta\cos\theta\partial_{\theta}V-2k^{2}\left(D-2-\sin^{2}\theta\right)V.
\end{equation}


\begin{thebibliography}{References}
\bibitem{Stephani} H.~Stephani, D.~Kramer, M.~MacCallum, C.~Hoenselaers,
and E.~Herlt, \emph{Exact Solutions of Einstein's Field Equations}
(Cambridge University Press, Cambridge, 2003).

\bibitem{Podolsky} J.~B.~Griffiths and J.~Podolský, \emph{Exact
Space-Times in Einstein's General Relativity} (Cambridge University
Press, Cambridge, 2009).

\bibitem{BHT-PRL} E.~Bergshoeff, O.~Hohm and P.~K.~Townsend,
\emph{{}``Massive Gravity in Three Dimensions,''} Phys.\ Rev.\ Lett.,
\textbf{102}, 201301 (2009).

\bibitem{LuPope} H.~Lu and C.~N.~Pope, \emph{{}``Critical Gravity
in Four Dimensions,''} Phys.\ Rev.\ Lett.\ \textbf{106}, 181302
(2011).

\bibitem{DeserLiu} S.~Deser, H.~Liu, H.~Lu, C.~N.~Pope, T.~C.~Sisman
and B.~Tekin, \emph{{}``Critical Points of D-Dimensional Extended
Gravities,''} Phys. Rev. \textbf{D83}, 061502 (2011).

\bibitem{Maldacena} J.~Maldacena, \emph{{}``Einstein Gravity from
Conformal Gravity,''} arXiv:1105.5632 {[}hep-th{]}.

\bibitem{Kundt} W.~Kundt, \emph{{}``The plane-fronted gravitational
waves,''} Zeitshrift für Physik \textbf{163}, 77, (1961).

\bibitem{ColeyHervik} A.~Coley, S.~Hervik, G.~O.~Papadopoulos
and N.~Pelavas, \emph{{}``Kundt Spacetimes,''} Class.\ Quant.\ Grav.\ \textbf{26},
105016 (2009).

\bibitem{DJT-PRL} S.~Deser, R.~Jackiw and S.~Templeton, \emph{{}``Three-Dimensional
Massive Gauge Theories,''} Phys.\ Rev.\ Lett.\ \textbf{48}, 975
(1982).

\bibitem{DJT-Annals} S.~Deser, R.~Jackiw and S.~Templeton, \emph{{}``Topologically
Massive Gauge Theories,''} Annals Phys.\ \textbf{140}, 372 (1982)
{[}Erratum-ibid.\ \textbf{185}, 406 (1988){]} {[}Annals Phys.\textbf{\ 185},
406 (1988){]} {[}Annals Phys.\ \textbf{281}, 409 (2000){]}.

\bibitem{Nutku} Y.~Nutku, \emph{{}``Exact solutions of topologically
massive gravity with a cosmological constant,''} Class.\ Quant.\ Grav.\ \textbf{10},
2657 (1993).

\bibitem{Gurses} M.~Gurses, \emph{{}``Perfect fluid sources in
2+1 dimensions,''} Class.\ Quant.\ Grav.\ \textbf{11}, 2585 (1994).

\bibitem{chakhad-2009} M.~Chakhad, \emph{{}``Kundt spacetimes of
massive gravity in three dimensions''}, arXiv:0907.1973 {[}hep-th{]};
PhD thesis.

\bibitem{Gurses-Killing} M.~Gurses, \emph{{}``Killing Vector Fields
in Three Dimensions: A Method to Solve Massive Gravity Field Equations,''}
Class.\ Quant.\ Grav.\ \textbf{27}, 205018 (2010) {[}Corrigendum
\textbf{29}, 059501 (2012){]}.

\bibitem{Aliev-PLB} H.~Ahmedov and A.~N.~Aliev, \emph{{}``The
General Type N Solution of New Massive Gravity,''} Phys.\ Lett.\ B
\textbf{694}, 143 (2010).

\bibitem{Aliev-PRL} H.~Ahmedov and A.~N.~Aliev, \emph{{}``Exact
Solutions in D-3 New Massive Gravity,''} Phys.\ Rev.\ Lett.\ \textbf{106},
021301 (2011).

\bibitem{Aliev-PRD} H.~Ahmedov and A.~N.~Aliev, \emph{{}``Type
D Solutions of 3D New Massive Gravity,''} Phys.\ Rev.\ D \textbf{83},
084032 (2011).

\bibitem{Chow-Classify} D.~D.~K.~Chow, C.~N.~Pope and E.~Sezgin,
\emph{{}``Classification of solutions in topologically massive gravity,''}
Class.\ Quant.\ Grav.\ \textbf{27}, 105001 (2010).

\bibitem{Chow-Kundt} D.~D.~K.~Chow, C.~N.~Pope and E.~Sezgin,
\emph{{}``Kundt spacetimes as solutions of topologically massive
gravity,''} Class.\ Quant.\ Grav.\textbf{\ 27}, 105002 (2010).

\bibitem{Siklos} S.~T.~C.~Siklos, \emph{{}``Lobatchevski plane
gravitational waves,''} in \emph{Galaxies, axisymmetric systems and
relativity}, ed.\ M.~A.~H.~MacCallum, (Cambridge University Press,
1985), 247\textendash{}274. 

\bibitem{Chamblin} A.~Chamblin and G.~W.~Gibbons, \emph{{}``Nonlinear
Supergravity on a Brane without Compactification,''} Phys.\ Rev.\ Lett.\ \textbf{84},
1090 (2000).

\bibitem{Gullu_Gurses} I.~Gullu, M.~Gurses, T.~C.~Sisman and
B.~Tekin, \emph{{}``AdS Waves as Exact Solutions to Quadratic Gravity,''}
Phys.\ Rev.\ D \textbf{83}, 084015 (2011).

\bibitem{Giribet} E.~Ayon-Beato, G.~Giribet and M.~Hassaine, \emph{{}``Bending
AdS Waves with New Massive Gravity,''} JHEP \textbf{0905}, 029 (2009).

\bibitem{Pravda} T.~Malek and V.~Pravda, \emph{{}``Type III and
N solutions to quadratic gravity,''} Phys.\ Rev.\ D \textbf{84},
024047 (2011).

\bibitem{Malek} T.~Malek, \emph{{}``Exact Solutions of General
Relativity and Quadratic Gravity in Arbitrary Dimension,''} arXiv:1204.0291
{[}gr-qc{]}; PhD thesis.

\bibitem{Alishah} M.~Alishahiha and R.~Fareghbal, \emph{{}``D-Dimensional
Log Gravity,''} Phys.\ Rev.\ D \textbf{83}, 084052 (2011).

\bibitem{coley-hervik-pelavas-2006} A.~Coley, S.~Hervik and N.~Pelavas,
\emph{{}``On spacetimes with constant scalar invariants,''} Class.\ Quant.\ Grav.\ \textbf{23},
3053-3074 (2006)..

\bibitem{Hervik} A.~A.~Coley, G.~W.~Gibbons, S.~Hervik and C.~N.~Pope,
\emph{{}``Metrics With Vanishing Quantum Corrections,''} Class.\ Quant.\ Grav.\ \textbf{25},
145017 (2008).

\bibitem{coley} A.~Coley, \emph{{}``Classification of the Weyl
tensor in higher dimension and applications''} Class.\ Quant.\ Grav.\ \textbf{25},
033001 (2008).

\bibitem{coley-hervik-pelavas-2008} A.~Coley, S.~Hervik and N.~Pelavas,
\emph{{}``Lorentzian spacetimes with constant curvature invariants
in three dimensions,''} Class.\ Quant.\ Grav.\ \textbf{25}, 025008(2008).

\bibitem{coley-hervik-pelavas-2009} A.~Coley, S.~Hervik and N.~Pelavas,
\emph{{}``Spacetimes characterized by their scalar curvature invariants,''}
Class.\ Quant.\ Grav.\ \textbf{26}, 025013(2009).

\bibitem{Fuster} A.~Fuster Perez, \emph{{}``Kundt spacetimes in
general relativity and supergravity,''} PhD thesis, Vrije University,
Amsterdam, (2007) (unpublished).

\bibitem{gurses-sisman-tekin-2011} M.~Gurses, T.~C.~Sisman and
B.~Tekin, \emph{{}``Some exact solutions of all $f(R_{\mu\nu})$
theories in three dimensions,''} arXiv:1112.6346 {[}hep-th{]}.

\bibitem{KerrSchild} R.~P.~Kerr and A.~Schild, \emph{{}``Some
algebraically degenerate solutions of Einstein's gravitational field
equations,''} Proc.\ Symp.\ Appl.\ Math.\textbf{\ 17 }199\textbf{
}(1965); G.~C.~Debney, R.~P.~Kerr and A.~Schild , \emph{{}``Solutions
of the Einstein and Einstein-Maxwell Equations,''} J.\ Math.\ Phys.\textbf{\ 10}
1842 (1969).

\bibitem{GursesGursey} M.~Gurses and F.~Gursey, \emph{{}``Lorentz
covariant treatment of the Kerr\textendash{}Schild geometry,''} J.\ Math.\ Phys.\ \textbf{16}
2385 (1975).

\bibitem{Anabalon} A.~Anabalon, N.~Deruelle, Y.~Morisawa, J.~Oliva,
M.~Sasaki, D.~Tempo and R.~Troncoso, \emph{{}``Kerr-Schild ansatz
in Einstein-Gauss-Bonnet gravity: An exact vacuum solution in five
dimensions,''} Class.\ Quant.\ Grav.\ \textbf{26}, 065002 (2009).

\bibitem{Pravda-KSwithAdS} T.~Malek and V.~Pravda, \emph{{}``Kerr-Schild
spacetimes with (A)dS background,''} Class.\ Quant.\ Grav.\ \textbf{28},
125011 (2011).

\bibitem{DeserTekin} S.~Deser and B.~Tekin, \emph{{}``Gravitational
energy in quadratic curvature gravities,''} Phys.\ Rev.\ Lett.\textbf{\ 89},
101101 (2002).

\bibitem{Deser_Tekin} S.~Deser, B.~Tekin, \emph{{}``Energy in
generic higher curvature gravity theories,''} Phys.\ Rev.\ \textbf{D67},
084009 (2003).

\bibitem{Gullu_Tekin} I.~Gullu and B.~Tekin, \emph{{}``Massive
Higher Derivative Gravity in D-dimensional Anti-de Sitter Spacetimes,''}
Phys.~Rev.~D \textbf{80}, 064033 (2009).

\bibitem{Higuchi} A.~Higuchi, \emph{{}``Symmetric Tensor Spherical
Harmonics On The N Sphere And Their Application To The De Sitter Group
${\rm So}(n,1)$,''} J.\ Math.\ Phys.\ \textbf{28}, 1553 (1987)
{[}Erratum-ibid.\ \textbf{43}, 6385 (2002){]}.

\bibitem{Bergshoeff-Log} E.~A.~Bergshoeff, O.~Hohm, J.~Rosseel
and P.~K.~Townsend, \emph{{}``Modes of Log Gravity,''} Phys.\ Rev.\ D
\textbf{83}, 104038 (2011).

\bibitem{Lu-LinModes} Y.~-X.~Chen, H.~Lu and K.~-N.~Shao, \emph{{}``Linearized
Modes in Extended and Critical Gravities,''} Class.\ Quant.\ Grav.\ \textbf{29},
085017 (2012).

\bibitem{Gurses_f(Riem)} M.~Gurses, T.~C.~Sisman and B.~Tekin,
\emph{{}``Kerr-Schild-Kundt type solutions of f(Riemann) theories''},
work in progress.

\bibitem{Hervik-BoostWeight} A.~Coley, S.~Hervik and N.~Pelavas,
\emph{{}``Lorentzian manifolds and scalar curvature invariants,''}
Class.\ Quant.\ Grav.\ \textbf{27}, 102001 (2010).

\bibitem{Chandra} S.~Chandrasekhar, \emph{The Mathematical Theory
of Black Holes} (Oxford University Press, New York, 1983).

\bibitem{Xanthopoulos} B.~C.~Xanthopoulos, \emph{{}``Exact vacuum
solutions of Einstein\textquoteright{}s equation from linearized solutions,''}
J.\ Math.\ Phys.\ \textbf{19}, 1607, 1978.

\bibitem{DereliGurses} T.~Dereli and M.~Gurses, \emph{{}``The
Generalized Kerr-schild Transform In Eleven-dimensional Supergravity,''}
Phys.\ Lett.\ \textbf{B171}, 209 (1986).\end{thebibliography}
\end{document}